\documentclass[]{fairmeta}

\usepackage{graphicx}
\usepackage{amsmath}
\usepackage[version=4]{mhchem}
\usepackage{graphicx} 
\usepackage{booktabs}
\usepackage{caption}
\usepackage{pifont}
\usepackage{soul}

\usepackage[utf8]{inputenc} 
\usepackage[T1]{fontenc}    
\usepackage{hyperref}       
\usepackage{url}            
\usepackage{booktabs}       
\usepackage{amsfonts}       
\usepackage{nicefrac}       
\usepackage{microtype}      
\usepackage{xifthen}
\usepackage{multirow} 
\usepackage{cleveref}
\usepackage{adjustbox}
\usepackage{mhchem}
\usepackage{threeparttable}
\usepackage{import}
\usepackage{epstopdf}
\usepackage{shellesc}
\usepackage{uniquecounter}

\newcommand{\oc}{OC25}

\newcommand{\angs}{\text{\AA}}

\newcommand{\kindatiny}{\fontsize{6pt}{7.2pt}\selectfont}
\newlength\savewidth
\newcommand{\tablestyle}[2]{%
    \fontfamily{ptm}\selectfont%
    \let\itold\it%
    \def\it{\itold \fontfamily{ptm}\selectfont}%
    \setlength{\tabcolsep}{#1}\renewcommand{\arraystretch}{#2}\centering\kindatiny%
    \let\citeold\cite%
    \renewcommand{\cite}[1]{\normalfont\fontfamily{ptm}\selectfont\tiny\citeold{##1}}%
}
\newcolumntype{x}[1]{>{\centering\arraybackslash}p{#1pt}}
\newcolumntype{y}[1]{>{\raggedright\arraybackslash}p{#1pt}}
\newcolumntype{z}[1]{>{\raggedleft\arraybackslash}p{#1pt}}
\newcolumntype{w}{>{\centering\arraybackslash}p{18pt}}
\newcolumntype{a}{>{\centering\arraybackslash}p{16pt}}

\definecolor{c0-title-bkg}{HTML}{ffffff}
\definecolor{c0-title-text}{HTML}{000000}
\definecolor{c0-item-bkg}{HTML}{ffffff}
\definecolor{c0-item-text}{HTML}{818589}

\definecolor{c1-title-bkg}{HTML}{d1e2dd}
\definecolor{c1-title-text}{HTML}{005953}
\definecolor{c1-item-bkg}{HTML}{e6efec}
\definecolor{c1-item-text}{HTML}{2d7b6d}

\definecolor{c2-title-bkg}{HTML}{cfe1e1}
\definecolor{c2-title-text}{HTML}{005760}
\definecolor{c2-item-bkg}{HTML}{e4eeed}
\definecolor{c2-item-text}{HTML}{24797b}

\definecolor{c3-title-bkg}{HTML}{cddfe5}
\definecolor{c3-title-text}{HTML}{00536b}
\definecolor{c3-item-bkg}{HTML}{e2ecef}
\definecolor{c3-item-text}{HTML}{287687}

\definecolor{c4-title-bkg}{HTML}{cedce8}
\definecolor{c4-title-text}{HTML}{124e74}
\definecolor{c4-item-bkg}{HTML}{e1eaf1}
\definecolor{c4-item-text}{HTML}{3a7190}

\definecolor{c5-title-bkg}{HTML}{d0d9eb}
\definecolor{c5-title-text}{HTML}{324779}
\definecolor{c5-item-bkg}{HTML}{e1e8f3}
\definecolor{c5-item-text}{HTML}{4d6b97}

\definecolor{c6-title-bkg}{HTML}{d3d5ed}
\definecolor{c6-title-text}{HTML}{493e7b}
\definecolor{c6-item-bkg}{HTML}{e3e5f5}
\definecolor{c6-item-text}{HTML}{61639b}

\definecolor{c7-title-bkg}{HTML}{dad1ed}
\definecolor{c7-title-text}{HTML}{5a3477}
\definecolor{c7-item-bkg}{HTML}{e5e1f5}
\definecolor{c7-item-text}{HTML}{725b99}

\definecolor{c8-title-bkg}{HTML}{ded1ec}
\definecolor{c8-title-text}{HTML}{633273}
\definecolor{c8-item-bkg}{HTML}{ebe2f6}
\definecolor{c8-item-text}{HTML}{7c5997}

\definecolor{c9-title-bkg}{HTML}{e5d1eb}
\definecolor{c9-title-text}{HTML}{6c2f6b}
\definecolor{c9-item-bkg}{HTML}{f0e0f6}
\definecolor{c9-item-text}{HTML}{885591}

\definecolor{c10-title-bkg}{HTML}{ebd1e7}
\definecolor{c10-title-text}{HTML}{722e5f}
\definecolor{c10-item-bkg}{HTML}{f5e2f3}
\definecolor{c10-item-text}{HTML}{915487}

\definecolor{avg-title-bkg}{HTML}{f3f3f3}
\definecolor{avg-title-text}{HTML}{000000}
\definecolor{avg-item-bkg}{HTML}{f3f3f3}
\definecolor{avg-item-text}{HTML}{000000}

\ExplSyntaxOn

\cs_generate_variant:Nn \coffin_typeset:Nnnnn {Nffff}

\NewDocumentCommand\rotbox{ O{l,H} D<>{0pt,0pt} m m}{
    \hcoffin_set:Nn \l_tmpa_coffin {#4}
    \coffin_rotate:Nn \l_tmpa_coffin {#3}
    \coffin_typeset:Nffff \l_tmpa_coffin 
        {\clist_item:nn{#1}{1}}
        {\clist_item:nn{#1}{2}}
        {\clist_item:nn{#2}{1}}
        {\clist_item:nn{#2}{2}}
}
\ExplSyntaxOff

\newlength{\ccustomlen}
\newcommand{\ccustom}[3][c0]{%
    \cellcolor{#1-item-bkg}{%
        \rotbox[l,t]{90}{%
            \parbox[t]{\ccustomlen}{%
                \ifthenelse{\isempty{#3}}{%
                    \mbox{%
                        \kindatiny\textcolor{#1-title-text}{#2}%
                    }%
                }{%
                    \kindatiny\textcolor{#1-title-text}{#2} \\%
                    \tiny{\textcolor{#1-item-text}{\it #3}}%
                }%
            }%
        }%
    }%
}

\usepackage{xspace}

\makeatother

\title{Insights into CO dimerization at electrified Cu interfaces from large-scale machine learning simulations}

\author[1]{Sushree Jagriti Sahoo}
\author[2]{Mikael Maraschin}
\author[3]{Joel B Varley}
\author[1]{Daniel S. Levine}
\author[1]{Zachary Ulissi}
\author[1]{C. Lawrence Zitnick}
\author[4]{Wayu Takemura}
\author[2,\dagger]{Joseph A. Gauthier}
\author[4\dagger]{Nitish Govindarajan}
\author[1,\dagger]{Muhammed Shuaibi}

\affiliation[1]{FAIR at Meta}
\affiliation[2]{Department of Chemical Engineering, Texas Tech University, Lubbock, TX 79409, USA}
\affiliation[3]{Materials Science Division, Lawrence Livermore National Laboratory, Livermore, CA 94550, USA}
\affiliation[4]{School of Chemistry, Chemical Engineering and Biotechnology, Nanyang Technological University, 21 Nanyang Link, Singapore 637371, Singapore}

\contribution[\dagger]{Co-corresponding Author}

\abstract{

Catalysis at solid--liquid interfaces underpins many energy technologies, yet \textit{ab initio} simulations that capture interfacial dynamics remain prohibitively expensive. Here we introduce Open Catalyst 2025 (OC25), the largest dataset for solid--liquid interfaces. To demonstrate OC25-trained models as practical tools for electrocatalysis, we investigate CO dimerization on Cu surfaces, a key step in CO$_2$ electroreduction. Using large cells (>800 atoms) and enhanced sampling up to 7 ns – the largest explicit-solvent CO dimerization study to date – we compute free-energy profiles under varied surface charge, cation identity, and surface facet. We find that dimerization is weakly sensitive to charge and cation identity, with appreciable stabilization only at the most negative charge densities, while extension to stepped Cu(310) reveals a more favorable pathway at modest reducing potentials. Our results demonstrate that OC25-trained models provide a scalable tool for investigating electrocatalytic transformations at solid--liquid interfaces, enabling simulations orders of magnitude beyond \textit{ab initio} methods.
}

\date{\today}
\correspondence{J.A.G.(\email{joe.gauthier@ttu.edu}), N.G.(\email{nitish.govindarajan@ntu.edu.sg}), M.S.(\email{mshuaibi@meta.com})}

\metadata[Code]{\url{https://github.com/facebookresearch/fairchem}}
\metadata[Dataset+Models]{\url{https://huggingface.co/facebook/OC25}}

\let\oldaddcontentsline\addcontentsline
\renewcommand{\addcontentsline}[3]{}
\usepackage{float}
\begin{document}
\maketitle

\section{Introduction}
\label{section:intro}

Solid-liquid interfaces are at the heart of several critical technologies, including catalysis, batteries, sensors, among others~\cite{harraz2025homogeneous,schiffer2017electrification,mallapragada2023decarbonization,barecka2023towards,cabana2022ngene,xia2022emerging,miao2023electrified,zhang2025electrolytic,chung2024direct,zhang2023advances,wang2023electrochemical,gouda2024green,wu2022energy,tao2022engineering,olusegun2024understanding,iriawan2021methods,lazouski2019understanding,fu2024calcium,goyal2024metal,wang2024fast,du2024side,liu2024revealing,bai2024low,zhang2024emerging,cai2024challenges,li2024recent,hao2025recent,singh2024review}. In contrast to gas-phase heterogeneous catalysis, electrocatalytic processes at solid-liquid interfaces are complicated by the presence of the liquid phase, where solvent and ion effects are intimately coupled with surface reactivity~\cite{gross2022abinitio,sundararaman2022improving,saleheen2023understanding,klemm2023impact,zhang2020method,kundu2025liquid, levell2024emerging,zhang2023promoting}. This electrochemical interface is inherently dynamic, with the electric double layer forming an integral part of the active site. Nevertheless, the interface is frequently approximated as a static surface that neglects the explicit electrolyte environment, primarily because of computational cost constraints~\cite{ringe2021implicit, gauthier2019unified}. Recognizing the complexity of electrochemical interfaces is particularly important for reactions such as CO$_2$ and CO electroreduction, where product selectivity depends sensitively on the catalyst structure, local pH, electrolyte identity, adsorbate coverage, and electrode potential~\cite{hahn2017engineering,resasco2025universal,kastlunger2022pH}.

\textit{Ab initio} molecular dynamics (AIMD) offers a rigorous, albeit computationally expensive, framework for simulating electrochemical interfaces~\cite{gross2022abinitio}. Due to its computational cost, AIMD simulations of metal/electrolyte interfaces are typically confined to modest cell sizes and short simulation timescales~\cite{gross2022abinitio, kristoffersen2021towards, heenen2020solvation, Monteiro2021}. These constraints make it difficult to obtain well-converged statistics for solvent reorganization, ion distributions, adsorbate rearrangements, and rare events, all of which are crucial for understanding electrochemical interfaces. Consequently, mechanistic investigations largely rely on static DFT, implicit solvent models (sometimes augmented with a few explicit molecules), or short-timescale MD simulations~\cite{ringe2021implicit,govindarajan2025intricacies,gross2022abinitio}. While such approaches have yielded valuable mechanistic insights, they may not fully capture how interface dynamics govern reaction pathways and catalytic performance~\cite{amirbeigiarab2023atomic,harraz2025homogeneous}.

Machine learning interatomic potentials (MLIPs) provide a compelling strategy to mitigate these constraints by enabling near-first-principles accuracy at substantially reduced cost~\cite{qm9,eSEN,eastman2023spice,deng2023chgnet,wood2025family,liao2023equiformerv2,batatia2023mace,gasteiger2022gemnet,batzner20223}. However, the most widely used large-scale catalyst datasets have traditionally emphasized solid--gas interfaces~\cite{oc20,oc22,aqcat25}, limiting their direct applicability to electrochemical environments where solvent, ions, and interfacial charge reshape both structure and reactivity. While recent efforts have begun to address this, they remain limited in size and chemical diversity~\cite{zhuang2025artificial, hormann2025machine}. To bridge this gap, we introduce the Open Catalyst 2025 (OC25) dataset and models for solid-liquid interfaces, comprising over 7 million DFT calculations sampling more than one million explicit solvent configurations across a wide variety of elements, solvents, ions, and non-equilibrium structures.

In the present work, we demonstrate that OC25-trained MLIPs can serve as practical scientific tools for atomistic simulations of electrocatalytic reactions. As a case study, we investigate CO dimerization on copper surfaces. Among metal electrocatalysts, copper is unique in producing appreciable amounts of multicarbon products, with CO dimerization widely regarded as a key step in the formation of C--C bonds in these pathways~\cite{hori2008electrochemical,nitopi2019progress,garza2018mechanism,goodpaster2016identification}. The energetics of this step are known to depend on applied potential, surface facet, CO coverage, local electric field, solvent structure, and cation identity~\cite{perez2017structure,resasco2017promoter}, a complexity that has proven difficult to resolve at the atomic scale.

We first validate the OC25 models against AIMD references of electrified metal-water interfaces, then use large-cell enhanced sampling simulations up to 7~ns to compute free-energy profiles for CO dimerization on copper surfaces under systematically varied surface charge, cation identity, and surface facet. We find that dimerization is largely insensitive to surface charge and cation identity, with appreciable stabilization emerging only at the most negative charge densities – a feature entirely absent from implicit-solvent predictions, while extension to stepped Cu(310) reveals a substantially larger facet effect that lowers both the barrier and reaction energy. Together, these results demonstrate that OC25-based MLIPs provide a scalable route to studying electrocatalytic reactions at electrified solid-liquid interfaces, enabling explicit-solvent simulations on length and timescales beyond the reach of \textit{ab initio} methods. The dataset, models, and code are all openly available.

\begin{figure}[t!]
    \centering
    \includegraphics[width=\linewidth]{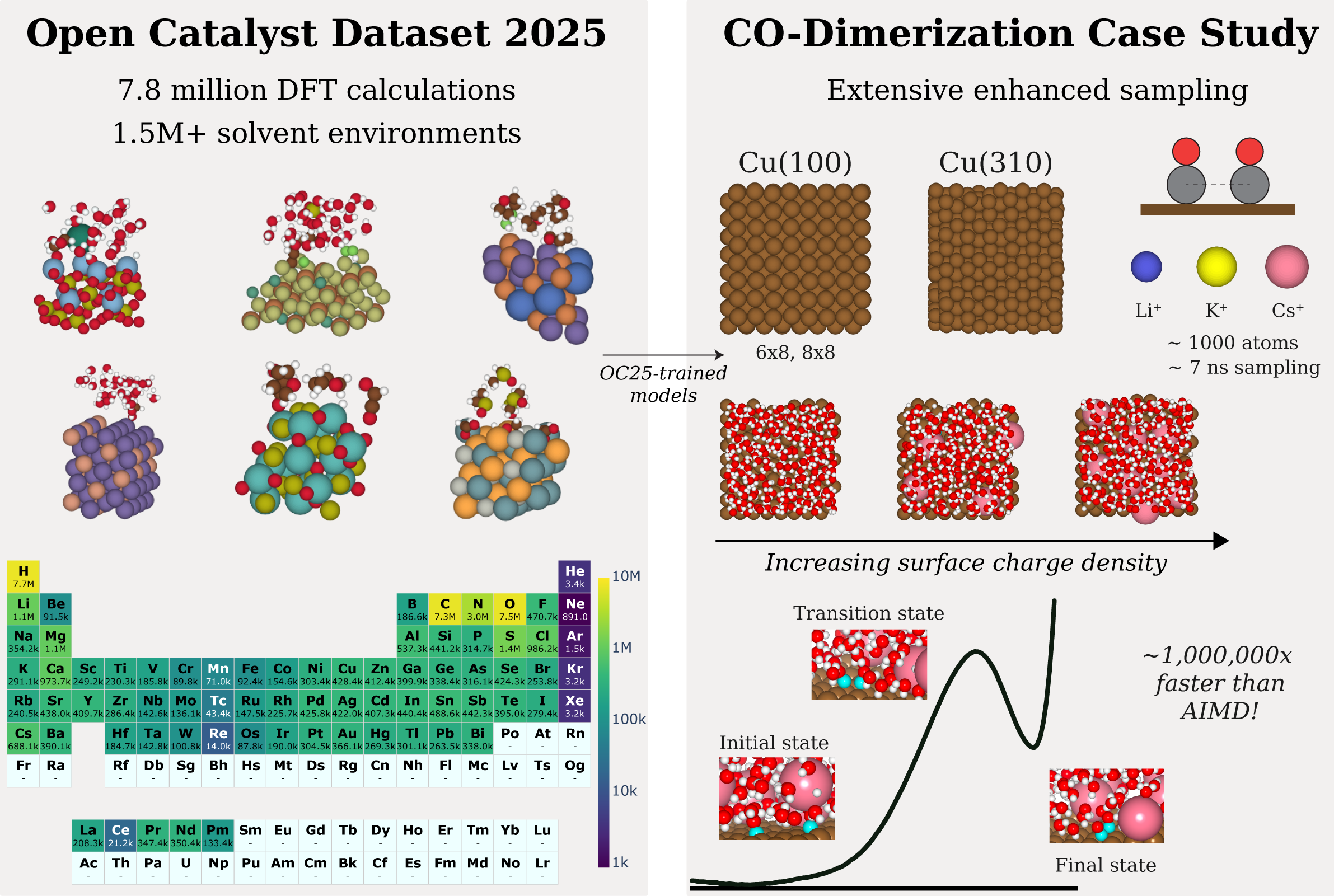}
    \caption{
    Overview of the Open Catalyst 2025 dataset and CO dimerization case study. Representative dataset snapshots and a periodic table indicating the number of training structures containing each element are shown. A trained OC25 model is then applied to the CO dimerization case study on Cu(100) and Cu(310) surfaces at varying surface charge densities and cation identities. A sample resulting free-energy profile and representative snapshots of the initial, transition, and final states are also provided.
    }
    \label{fig:overview}
\end{figure}
\section{Results}\label{section:results}

\subsection{OC25 Dataset}

The Open Catalyst 2020/2022 datasets (OC20/OC22)~\cite{oc20,oc22}, Catalysis-Hub~\cite{winther}, GASpy~\cite{tran2018active}, and AQCat~\cite{aqcat25} represent the largest catalyst datasets for training MLIPs. The diversity and scale of these datasets have led to the development of several state-of-the-art ML models for the community~\cite{liao2023equiformerv2,gasteiger2022gemnet,eSEN}. Although models trained on these datasets have demonstrated successful applications for gas-phase heterogeneous catalysis, little has been done for solid-liquid and electrified solid-liquid interfaces~\cite{lan2023adsorbml, wander2025cattsunami, moon2025catbench}. The Open Catalyst 2025 (OC25) dataset aims to bridge this gap by presenting the largest catalyst solid-liquid interface dataset.

The OC25 dataset is constructed from a combination of DFT relaxations and AIMD simulations of catalyst, solvent, and ion structures. Structures consist of a catalyst surface, a solvent, and at least one adsorbate molecule. Structures may also contain multiple adsorbates and an ion to better capture reactive environments and electrocatalytic systems. Surfaces are sampled from unique bulk structures in the Materials Project~\cite{jain2020materials}, including oxides. Eight commonly used solvents are sampled across varying solvent depths, with system sizes of 144 atoms on average. Adsorbates are randomly sampled from 98 unique molecules, including OC20 species and additional reactive intermediates. A more detailed description of the dataset generation can be found in the Methods section. Overall, the OC25 dataset consists of 7,801,261 single-point calculations, spanning 1,511,270 unique systems and 88 unique elements (Figure \ref{fig:overview}).  

OC25 structures correspond to highly off-equilibrium configurations, a property that we have seen to aid in training ML models~\cite{omat24, oc20, eSEN}. To accomplish this, short-timescale AIMD simulations at high temperature (1000K) were run, minimizing the redundancy in structures that can come from fully relaxed configurations, i.e. OC20 and OC22. This is reflected in the force distributions in Supplementary Figure~\ref{fig:oc25_comp_dist}, with higher forces than those of OC20 and OC22.

\begin{figure}[ht!]
    \centering
    \includegraphics[width=0.90\textwidth]{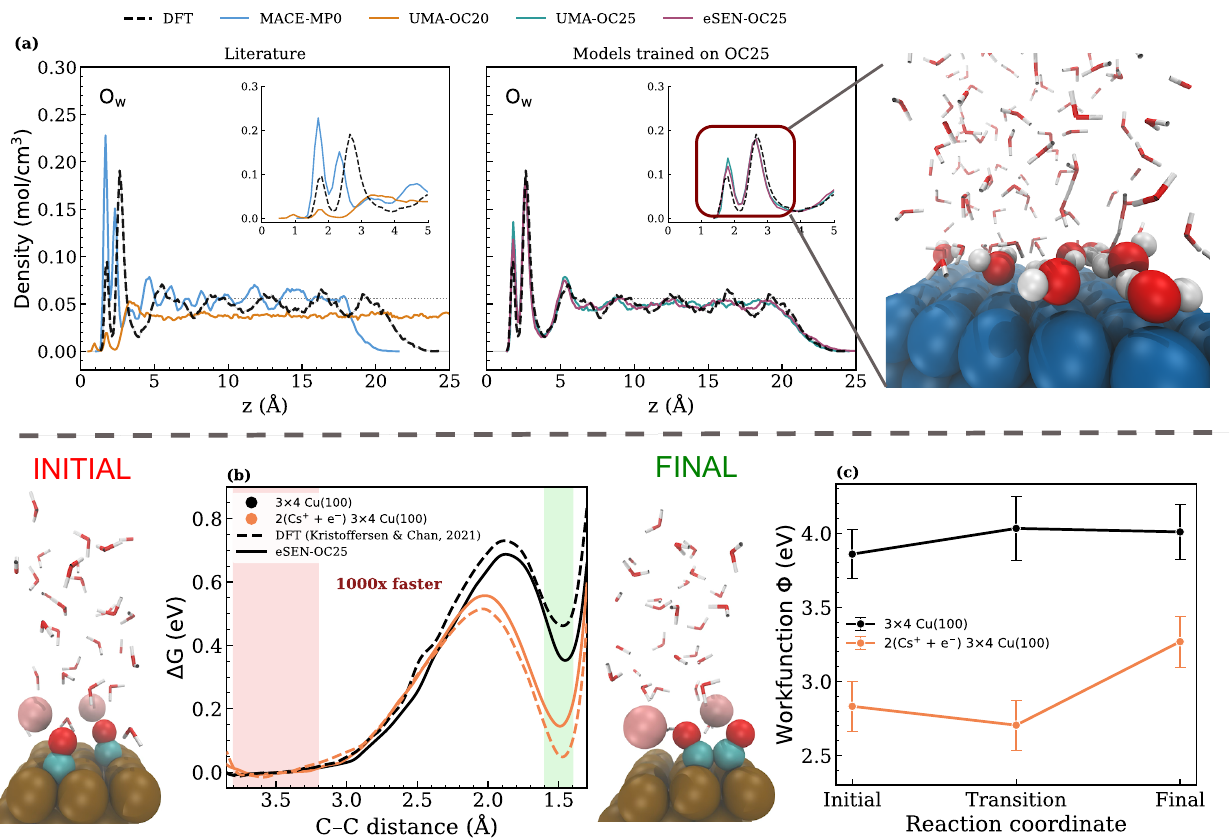}
    \vspace{10pt}
    \caption{Validation of OC25-trained models against literature references. \textbf{(a)} Oxygen density profiles ($\text{O}_\text{w}$) along the surface normal $z$ for previous-generation and OC25-trained universal models relative to the DFT reference~\cite{dominguez2024metal}. The inset highlights the chemisorbed water peak  at $z \approx 2$~\AA{}. The dotted horizontal line indicates the bulk water density. A close-up of the chemisorbed water layer at the Rh(111) surface as computed by eSEN-OC25 is also shown. \textbf{(b)} Free energy profile for CO dimerization as a function  of C--C distance on a $3\times4$ Cu(100)/water interface, comparing eSEN-OC25(solid) against AIMD reference data~\cite{kristoffersen2021towards}(dashed) at neutral (black) and 2~(Cs$^+$~+~e$^-$) imposing a cathodic potential (orange). Shaded regions indicate the 2 *CO initial state (red) and *OCCO dimer final state (green). \textbf{(c)} Average workfunctions of the initial, transition, and final state from $\sim$50 random single-point DFT calculations from each sampled state of the OC25 MD trajectories.}
    \label{fig:reproduce-lit}
\end{figure}

For this work, we train an eSEN~\cite{eSEN} energy conserving model trained on the OC25 dataset. eSEN represents a state-of-the-art graph neural network (GNN) model that operates on graphs where atoms are nodes and edges are the interactions between them. While models of various sizes are trained, we use the smaller (eSEN-S) model, striking a balance between accuracy and efficiency. We also note that UMA-1.2~\cite{wood2025family} has recently incorporated OC25 into its training mixture; the resulting UMA-OC25 model is included in our interfacial structure benchmarks but was not available at the time the enhanced sampling simulations in this work were performed. A detailed discussion on baseline models, training protocols, and benchmark results can be found in the Methods section.

\begin{figure}[ht!]
    \centering
    \includegraphics[width=0.90\textwidth]{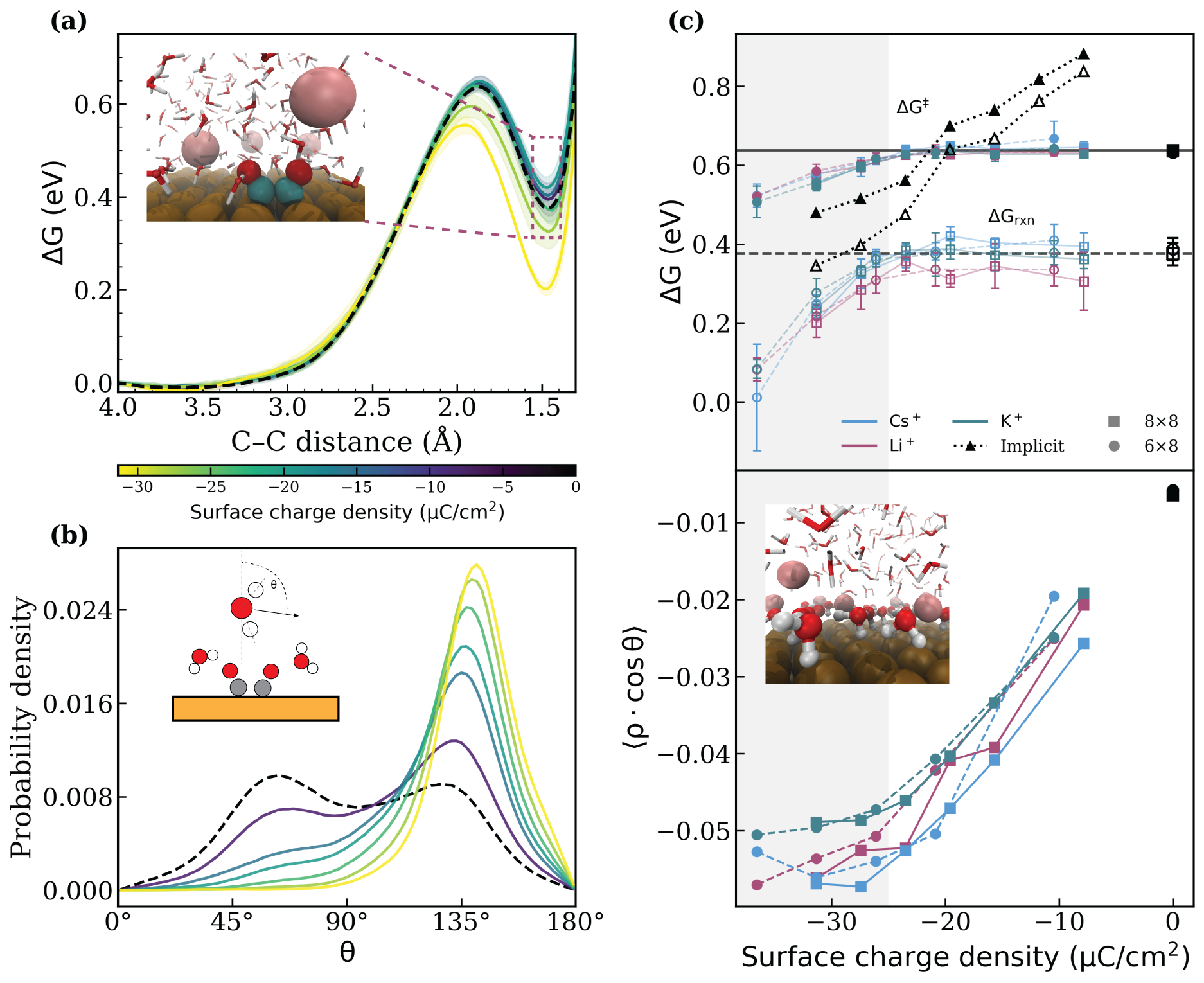}
    \caption{Effect of surface charging and cation identity on CO dimerization on Cu(100). \textbf{(a)} Free-energy profiles along the C--C distance collective variable as a function of interfacial Cs$^{+}$ concentration. The inset shows a representative *OCCO dimer configuration on Cu(100). \textbf{(b)} Water orientation distribution in the inner Helmholtz layer ($\sim$4.5\AA{}) as a function of interfacial Cs$^{+}$ concentration. \textbf{(c)} (top) Extracted activation barriers (filled) and reaction energies (open) from explicitly solvated enhanced sampling across various cation species (Cs$^+$, K$^+$, and Li$^+$) and cell sizes (8$\times$8 and 6$\times$8). The implicitly solvated harmonic oscillator transition state theory (hTST) approximation is also included (triangles). (bottom) Density-weighted orientations as a function of surface charge density, with the inset figure corresponding to the very negative ($-31$ $\mu$C/cm$^2$) 8 Cs$^+$ configuration. The shaded region highlights the most negative charge densities where appreciable stabilization of the dimerization barrier emerges.}
    \label{fig:dimerization_surf_charge}
\end{figure}

\subsection{OC25 models capture interfacial structure and reaction kinetics at solid-liquid interfaces where previous universal potentials fail}

We first benchmark the OC25-trained models against AIMD references of solid-liquid interfaces reported by Dom\'inguez-Flores et al.~\cite{dominguez2024metal}. The systems comprise five-layer $6\times6$ supercells of Rh(111), Pd(111), Pt(111), Ag(111), Au(111), and Ru(0001) with 144 explicit water molecules, simulated at 298~K for approximately 100~ps (details in Methods).

Among the six surfaces, Rh(111)/water is chosen as a representative case because its characteristic chemisorbed water layer presents a particularly challenging test for the MLIPs. As shown in Figure~\ref{fig:reproduce-lit}(a), foundation models such as MACE-MP-0 and UMA-OC20 fail to qualitatively reproduce the oxygen density profile at the interface obtained from AIMD: MACE-MP-0 predicts an unphysical spike at $z \approx 1$~\AA, while UMA-OC20 underestimates the near-surface water density and misses the layered structure entirely. In contrast, both OC25-trained models (eSEN-OC25 and UMA-OC25) quantitatively reproduce the AIMD density profile, including the chemisorbed water peak, solvation layers, and bulk density. Similar agreement is observed across all other metal interfaces (Supplementary Figures~\ref{fig:si:Ag111}--\ref{fig:si:Ru0001}).

To validate beyond interfacial structure, we evaluate reaction energetics via metadynamics simulations of CO dimerization on a 3$\times$4 Cu(100)/water interface, closely matching the settings of Kristoffersen and Chan~\cite{kristoffersen2021towards} (simulation details in Methods). We also probe double-layer charging by introducing two Cs$^+$ ions, imposing a surface charge density of $-41.8~\mu$C/cm$^2$. The resulting free-energy profiles (Figure~\ref{fig:reproduce-lit}(b)) show that both the AIMD reference and eSEN-OC25 exhibit a consistent reduction in the dimerization barrier with increasing negative surface charge, confirming that eSEN-OC25 captures the potential-dependent energetics. Differences of $\sim$0.1--0.15~eV between our results and the AIMD reference are consistent with the omission of dipole corrections detailed by Kristoffersen and Chan~\cite{kristoffersen2021towards}. In contrast, UMA-OC20 fails to reproduce the AIMD reference (Supplementary Figure~\ref{fig:si:Ru0001}), demonstrating that explicit training on solid-liquid interface data is essential for reproducing reaction energetics at electrified interfaces.

Work functions, $\Phi$, computed from $\sim$50 DFT single-points sampled along the eSEN-OC25 trajectories (Figure~\ref{fig:reproduce-lit}(c)) remain unchanged across the reaction coordinate in the absence of ions, but shift by $\sim$0.5~eV upon Cs$^+$ addition, consistent with observations of Kristoffersen and Chan~\cite{kristoffersen2021towards}. This highlights that small unit cells push the interface off the targeted potential under constant-charge conditions. Approaching a true constant-potential regime demands substantially larger interfacial areas, now within computational reach with OC25-trained models and explored in the following section.

\subsection{CO dimerization energetics are largely insensitive to surface charge except at very negative charge densities}

Having validated the OC25 models against AIMD benchmark studies, we leverage the extended length and timescales accessible with OC25-driven MD to study CO dimerization on Cu(100) and its dependence on surface charge. Simulations were performed on (6$\times$8) and (8$\times$8) Cu(100) surfaces using OPES~\cite{invernizzi2021opes} for approximately 7~ns (details in Methods), yielding well-converged free-energy profiles across a systematic variation of interfacial Cs$^+$ concentration that would be intractable using AIMD. Figure~\ref{fig:dimerization_surf_charge}(a) shows that across most of the sampled range, both the dimerization barrier ($\Delta G^{\ddagger}$) and reaction energy ($\Delta G_{\mathrm{rxn}}$) change only modestly. Appreciable stabilization emerges only at the highest Cs$^+$ concentrations (7--8 ions), corresponding to surface charge densities beyond $-25~\mu$C/cm$^2$.

This weak charge dependence is consistent with Kristoffersen and Chan~\cite{kristoffersen2021towards}, who attributed it to the dynamic reorganization of explicit interfacial water, which screens the surface charge along the *OCCO formation coordinate. Figure~\ref{fig:dimerization_surf_charge}(b) confirms that substantial water reorganization accompanies increasing surface charge in our simulations: with increasing charge density, interfacial water reorients from a bimodal distribution at the PZC to a single peak at ca.\ 135\textdegree{} with protons pointing toward the surface.

Figure~\ref{fig:dimerization_surf_charge}(c) compares the explicit-solvent results with implicit solvation and harmonic transition-state theory (hTST)~\cite{vineyard1957frequency}. In the most negative surface charge regime, hTST agrees quantitatively (within 0.1~eV) with the explicit-solvent results. However, the two diverge at less negative charges, where explicit water rearrangement compensates for the field-dipole stabilization of the transition state. The density-weighted water orientation shifts linearly with charge to approximately $-25~\mu$C/cm$^2$, beyond which it saturates. We posit that the agreement between hTST and explicit-solvent results at high charge densities reflects error cancellation, since linear charging models are strictly valid only near the potential of zero charge. This highlights that continuum solvation models cannot capture field-driven water reorientation effects that fundamentally alter the charge dependence of the dimerization energetics.

The large 8$\times$8 unit cell also addresses an important methodological limitation. Prior AIMD simulations on the small 3$\times$4 cells required post hoc constant-potential corrections~\cite{kristoffersen2021towards}. In contrast, the increased interfacial area in the present simulations significantly reduces reaction-induced work-function shifts (Table~\ref{tab:workfunction}), yielding approximately constant-potential free-energy profiles without correction. Both $\Delta G^{\ddagger}$ and $\Delta G_{\mathrm{rxn}}$ decrease with decreasing work function (Supplementary Figure~\ref{fig:si:wf_dependence}), with the reaction energy showing stronger potential dependence.

To validate the reliability of the eSEN-OC25 model in this out-of-distribution regime (larger systems, multiple ions, more complex interactions), DFT single-points were computed on $\sim$100 configurations per simulation sampled from the initial, transition, and final states. Across all ion species and counts, $\Delta G^{\ddagger}$ and $\Delta G_{\mathrm{rxn}}$ errors remain below 0.1~eV and force MAEs remain below 0.005~eV/\AA{} (Table~\ref{tab:ml_vs_dft}), consistent with training validation errors. Details are provided in Methods.

\begin{table}[t]
\centering
\begin{tabular}{@{}lcccc@{}}
\toprule
Ion & Count
& \shortstack{$\Delta G^{\ddagger}$ (meV)}
& \shortstack{$\Delta G_{\mathrm{rxn}}$ (meV)}
& \shortstack{Forces (meV/Å)} \\ \midrule
-                   & 0       & 46.8         & 92.6             & 4.06               \\ \midrule
\multirow{3}{*}{Cs} & 2       & 57.5         & 99.5             & 3.83               \\
                    & 5       & 43.1         & 85.5             & 3.61               \\
                    & 8       & 41.4         & 34.3             & 4.38               \\ \midrule
\multirow{3}{*}{K}  & 2       & 56.1         & 78.6             & 3.99               \\
                    & 5       & 46.2         & 86.8             & 4.03               \\
                    & 8       & 47.1         & 47.8             & 4.82               \\ \midrule
\multirow{3}{*}{Li} & 2       & 56.6         & 97.6             & 3.99               \\
                    & 5       & 36.3         & 82.5             & 3.98               \\
                    & 8       & 31.1         & 47.4             & 4.61               \\ \bottomrule
\end{tabular}
\caption{DFT validation of eSEN-OC25 predictions along the CO dimerization reaction coordinate on $8\times8$ Cu(100). $\sim$100 configurations are randomly sampled from each state (initial, transition, final) for single-point DFT evaluation. $\Delta G^{\ddagger}$ and $\Delta G_{\mathrm{rxn}}$ errors are computed from the mean energy of each state; force errors are reported as the mean absolute error (MAE) across all sampled configurations.}
\label{tab:ml_vs_dft}
\end{table}

\subsection{Cation identity has a minor impact on CO dimerization energetics}

Alkali cations are known to influence CO$_2$/CO reduction activity and selectivity, but the molecular origin of these effects remains debated~\cite{Xu2026,resasco2017promoter,ringe2019understanding,resasco2025universal}. To isolate the role of cation identity, we performed equivalent OPES simulations with Cs$^+$, K$^+$, and Li$^+$ at comparable ion concentrations. As shown in Figure~\ref{fig:dimerization_surf_charge}(c), cation-dependent differences in $\Delta G^{\ddagger}$ and $\Delta G_{\mathrm{rxn}}$ are small, at most 0.025 and 0.046~eV at a surface charge density of -31.3$~\mu$C/cm$^2$, within the statistical uncertainty of the sampling. By comparison, varying surface charge density from 0 to -31.3$~\mu$C/cm$^2$ shifts the barrier and reaction energies by 0.082 and 0.173~eV, respectively, indicating that the electrostatic state of the interface is a substantially stronger descriptor of the dimerization energetics than the cation identity.

Experimentally, the C$_2$ partial current densities vary by $\sim$2--10$\times$ across alkali cations on Cu(100)~\cite{resasco2017promoter}, corresponding to $\Delta\Delta G \approx k_\mathrm{B}T\ln(j_2/j_1) \approx 20\text{--}0.06~\mathrm{eV}$. Our computed barriers do not show a statistically significant cation ordering at this scale. While we cannot rule out subtle ion-specific effects on the *OCCO transition state, the influence of surface charge density on the dimerization energetics is substantially larger and more clearly resolved. This suggests that cation effects on C$_2$ selectivity likely arise from other factors, such as differences in packing density near the interface, local transport, CO availability, or competing elementary steps, not captured by the dimerization barrier alone. This interpretation is consistent with Morales-Guio et al.~\cite{morales2024electrochemical}, who argued that apparent cation-dependent selectivity is strongly influenced by mesoscopic transport, reactor geometry, and interfacial chemistry rather than intrinsic stabilization of C--C coupling intermediates.

\subsection{Nanosecond timescale simulations are required for converged CO dimerization energetics}

\begin{figure}[t]
    \centering
    \includegraphics[width=0.95\textwidth]{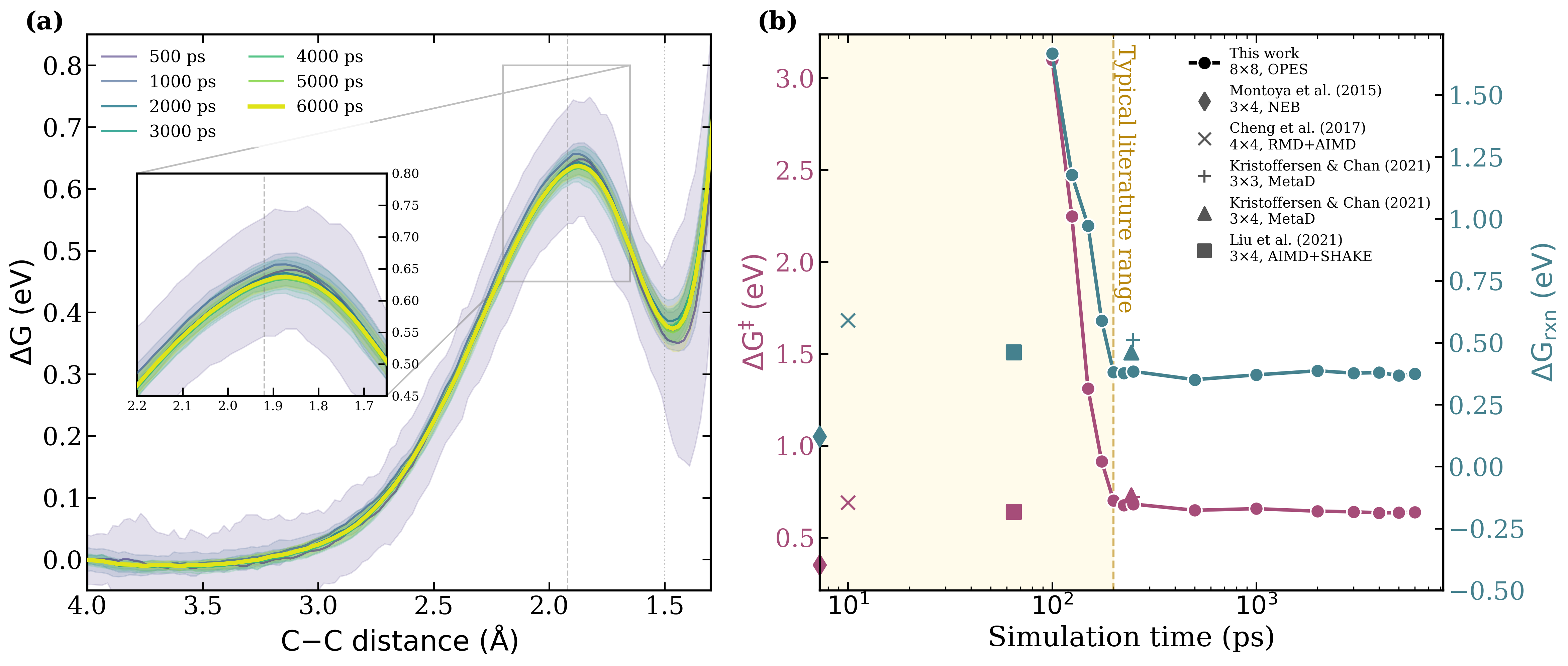}
    \caption{Sampling convergence of the CO dimerization on Cu(100). \textbf{(a)} Time-resolved free-energy profiles along the C--C distance from OPES with simulation windows from 500 ps to 6 ns; shaded bands are block uncertainties. \textbf{(b)} Convergence of $\Delta G^{\ddagger}$ and $\Delta G_{\mathrm{rxn}}$ versus simulation time. Circles are this work; other markers are literature values, with shape identifying the study. The shaded region ($\leq$200 ps) marks the typical timescale of prior studies.}
    \label{fig:sampling}
\end{figure}

The calculated free-energy profile for CO dimerization on Cu(100) exhibits a strong dependence on sampling time. As shown in Fig.~\ref{fig:sampling}(a), both the transition-state region ($d_{C-C}\approx$ 2\AA{}) and product basin ($d_{C-C}\approx$ 1.5\AA) are poorly converged within the first few hundred picoseconds. The profiles converge with increasing simulation time and after 500--1000~ps, the barrier height and reaction free energy become independent of the sampling duration. This indicates that CO dimerization is coupled to slow interfacial degrees of freedom - solvent reorganization, cation rearrangement, and *CO diffusion, which are insufficiently sampled on short timescales.

Figure~\ref{fig:sampling}(b) places our results in the context of prior AIMD studies~\cite{montoya2015theoretical, kristoffersen2021towards, cheng2017full,liu2021promotional}. Literature values exhibit a wide spread in $\Delta G^{\ddagger}$ and $\Delta G_{\mathrm{rxn}}$. These earlier simulations employed smaller unit cells (3$\times$3 to 4$\times$4) and substantially shorter sampling durations, typically 10–100 ps. Despite these constraints, several studies obtained barriers (0.64-0.73~eV) in agreement with our converged value ($\sim$0.64~eV), while reaction energies (0.12-0.59~eV) show wider spread around our converged estimate ($\sim$0.37~eV). The closer agreement in barriers likely reflects that well-designed enhanced sampling protocols can partially compensate for limited trajectory length, as demonstrated by the constrained AIMD approach of Liu et al.~\cite{liu2021promotional}. The larger scatter in reaction energies may reflect incomplete sampling of solvent configurations around the *OCCO product state. As a chemical step that is well-described by a single collective variable, CO dimerization represents a best-case scenario for enhanced sampling. More complex reactions involving charge transfer may require substantially longer sampling to account for electrolyte reorganization timescales.

Nevertheless, the computational cost of AIMD fundamentally limits the scope of accessible investigations. A single 7~ns OPES trajectory on our 8$\times$8 Cu(100) system completes in 12 days on a single NVIDIA H100 GPU, approximately $10^6\times$ faster than an equivalent DFT simulation, a speedup that persists regardless of the enhanced sampling algorithm employed. This throughput enables a qualitatively different mode of investigation: the systematic variation of surface charge, cation identity, and cell size in this work represents dozens of multi-nanosecond trajectories, a campaign entirely intractable at the DFT level. 

\subsection{Cu(310) exhibits a more favorable CO dimerization pathway compared to Cu(100)}

\begin{figure}[t!]
    \centering
    \includegraphics[width=0.9\textwidth, trim={0 32 0 0}, clip]{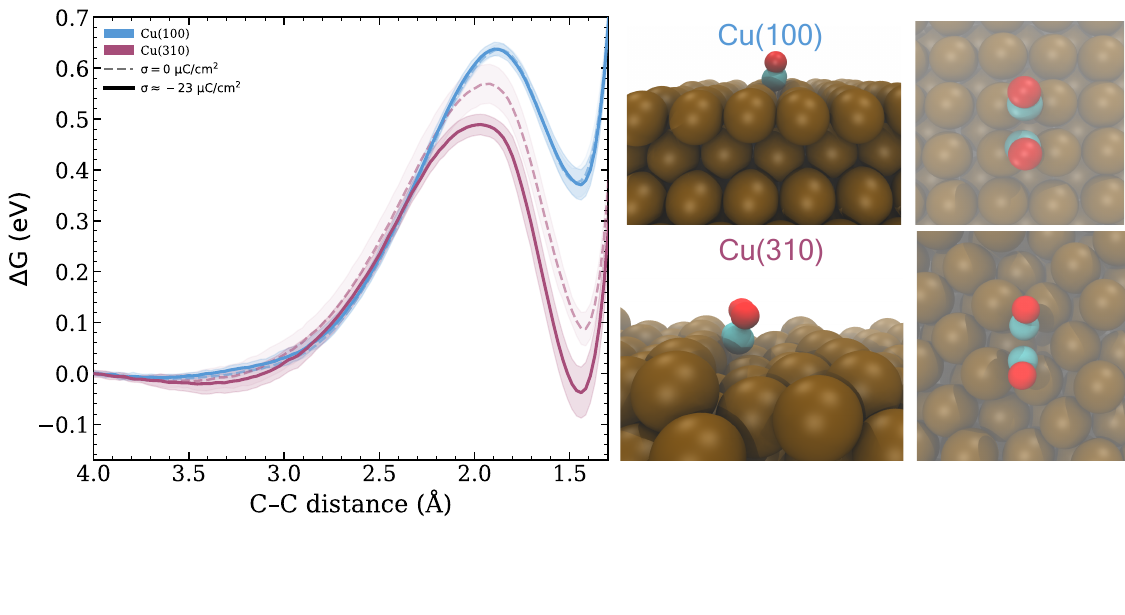}
    \caption{Effect of surface charging on CO dimerization on Cu(310) vs Cu(100) at the same surface charge density. Free energy profiles along the C--C distance collective variable between Cu(100) and Cu(310) for neutral and a charged surface at a $\sigma$ = -23 $\mu$C/cm$^2$. Representative top and side views of the Cu(100) and Cu(310) interfaces are shown for reference.
    }
    \label{fig:dimerization_facets}
\end{figure}

Across all of our simulations of CO dimerization on Cu(100), the reaction remains endergonic, in agreement with prior studies~\cite{kastlunger2022pH,kristoffersen2021towards,montoya2015theoretical,liu2021promotional,cheng2017full}. Although Cu(100) has been widely studied for C--C coupling during CO$_2$ reduction, calculated thermodynamics consistently predict uphill dimerization, contradicting the experimental observation of C$_{2+}$ products. Recent single-crystal experiments by Nguyen et al.~\cite{nguyen2024influence} suggest that C--C coupling does not occur on close-packed surfaces. They observed that both Cu(111) and Cu(100) initially exhibit near-unity Faradaic efficiency for hydrogen evolution, with significant C$_{2+}$ formation emerging only after several potential sweeps and surface reconstruction. These observations suggest that the active sites for C--C coupling may differ from the idealized terraces studied computationally.

Motivated by the unfavorable thermodynamics on Cu(100), we examine CO dimerization on Cu(310), where under-coordinated step-edge sites may provide a more favorable pathway for C–C coupling. As shown in Figure~\ref{fig:dimerization_facets}, the free-energy profiles exhibit a slightly lower barrier than on Cu(100). The barrier for Cu(310) is 0.57~eV at neutral conditions and 0.49~eV at $-23~\mu$C/cm$^2$, compared with 0.64~eV for Cu(100) at the corresponding conditions. The contrast in reaction energy is more pronounced. $\Delta G_{\mathrm{rxn}}$ for Cu(310) is 0.088~eV versus 0.375~eV for Cu(100) at neutral conditions, and becomes exergonic ($-0.037$~eV) at $-23~\mu$C/cm$^2$ while remaining strongly uphill on Cu(100) (0.371~eV). The dimer product basin is substantially more stable on the stepped facet, both with and without cations.

We extended the OPES simulations on Cu(310) to K$^+$ and Li$^+$. As shown in Supplementary Figure~\ref{fig:si:Cu310_barriers_energies}, both $\Delta G^{\ddagger}$ and $\Delta G_{\mathrm{rxn}}$  are governed almost entirely by surface charge density and are generally insensitive to cation identity, consistent with Cu(100). These results indicate that CO dimerization on Cu(310) is mildly exergonic even at modestly reducing potentials, supporting the experimental observation that stepped sites might be the active sites for C--C coupling towards C$_{2+}$ products~\cite{scholten2021identifying,gauthier2021role}.
\section{Discussion}\label{section:discussion}

The simulations presented here demonstrate that OC25-based MLIPs enable a qualitatively different mode of investigation for electrocatalytic reactions at solid–liquid interfaces. By providing near-DFT accuracy at a fraction of the cost, these models allow systematic exploration of interfacial conditions including surface charge, cation identity, and surface facet with explicit solvent on extended length and timescales. The large simulation cells used here (up to 8$\times$8 supercells) minimize reaction-induced work-function shifts, while nanosecond trajectories ensure adequate sampling of slow interfacial degrees of freedom together yielding approximately constant-potential, well converged free-energy profiles.

A central insight from this work is that the CO dimerization barrier on Cu(100) is largely insensitive to surface charge over much of the sampled potential window, with appreciable stabilization emerging only at the most negative charge densities. This weak charge dependence arises from the dynamic reorganization of interfacial water, which  screens the surface electric field and offsets much of the expected field–dipole stabilization of the *OCCO transition state. By comparison, implicit-solvent hTST models, which lack this solvent response, predict a substantially steeper charge dependence in the low-charge regime, which is a qualitative difference that highlights the importance of explicit representations of the electrolyte. Meanwhile, cation identity exerts only a minor influence on the dimerization energetics compared to surface charge or facet; any cation-dependent differences fall within the statistical uncertainty of our simulations. The extension to Cu(310) further illustrates how readily OC25-driven simulations can test mechanistic hypotheses: the substantially more favorable dimerization on the stepped surface suggests that undercoordinated sites may play a more important role in C--C coupling than the widely studied flat terraces.

There are several natural extensions to the present work. The constant-charge simulations used here approximate but do not rigorously enforce constant-potential conditions; coupling OC25 models with grand-canonical approaches or developing models that directly predict the work function would strengthen future studies. Additionally, the model architectures employed here rely on local message passing with finite cutoffs and thus do not explicitly capture long-range electrostatics that govern ion–ion and ion–surface interactions. While this did not appear to limit accuracy for the systems studied here, it may become more consequential for systems with stronger electrostatic ordering, motivating the development of architectures that more explicitly encode long-range interactions.

The simulations demonstrated here are made possible by the scale and diversity of the OC25 dataset. Previous MLIPs for catalysis either lacked training data for solid–liquid interfaces entirely or required system-specific fine-tuning to achieve acceptable accuracy for a single surface/solvent combination. In contrast, the breadth of OC25 yields a general-purpose potential that transfers directly to the systems studied here without additional training. This general transferability unlocks the length scales, timescales, and systematic studies demonstrated in this work. While CO dimerization on Cu(100) converges within a few nanoseconds, more complex chemistries–involving multi-step reaction networks, larger adsorbates, restructuring surfaces, or slower solvent dynamics–will likely require substantially longer sampling. The ability to routinely access systems of 1,000+ atoms on timescales of nanoseconds and beyond with MLIPs ensures that these more demanding problems remain tractable, and highlights the importance of continued development of both models and sampling methodologies for electrocatalytic applications. We hope that the openly available OC25 dataset and models will enable the community to explore increasingly complex and diverse interfacial chemistries beyond what has previously been accessible.

\section{Acknowledgements}
N. G. acknowledges support from a startup grant at NTU (award number 024462-00001).

M.M. and J.A.G. gratefully acknowledge support from The Welch Foundation under Grant Number D-2188-20240404.

J.B.V. acknowledges support by the U.S. Department of Energy, Lawrence Livermore National Laboratory (LLNL) under Contract No. DE-AC52-07NA27344.
\section{Methods}\label{sec:methods}
\subsection{OC25 Dataset Generation}\label{section:dataset}
The OC25 dataset is generated following a similar pipeline to that of OC20~\cite{oc20}, with the addition of a solvated interface. Structures were created in three stages: (1) adsorbate+surface generation, (2) interface construction, and (3) \textit{ab initio} calculations. All code to generate configurations is provided at \href{https://github.com/facebookresearch/fairchem/blob/main/src/fairchem/data/oc/core/interface_config.py}{https://github.com/facebookresearch/fairchem/}.

\subsubsection{Adsorbate+Surface generation}
We begin by constructing adsorbate+surface structures in vacuum. A bulk material is randomly sampled from a set of 39,821 materials in the Materials Project. For the sampled bulk material, all symmetrically distinct surfaces with Miller indices less than or equal to 3 are enumerated and a random surface is selected. The surface is tiled in the xy-plane to a length of 8\angs. The number of adsorbates per structure is randomly sampled between 1 and 5, with a single adsorbate selected 50\% of the time. Of the configurations with multiple adsorbates, 20\% consist of identical molecules. The adsorbates are sampled from the original set of OC20 adsorbates, containing oxygen, hydrogen, C1/C2 molecules, and nitrogen-containing species \cite{oc20}. This set is extended to also include reactive intermediates from the OC20NEB~\cite{wander2025cattsunami} and OCx24~\cite{abed2024open} (Supplementary Table \ref{tab:adsorbates}). Adsorbate(s) placement is performed using the \href{https://github.com/facebookresearch/fairchem/blob/dba2f17b46273533eccd7768a8a0fd028878fe86/src/fairchem/data/oc/core/adsorbate_slab_config.py#L63-L67}{Adsorb-ML workflow}~\cite{lan2023adsorbml}. Adsorbates are randomly placed on sites selected from Delaunay triangulation of surface atoms, followed by rotations along the z-axis and wobbles around the x/y-axis. In structures with multiple adsorbates, sites are only considered if their distance to the nearest adsorbate does not result in considerable overlap ($r_{cov} + 0.1$\angs).

\subsubsection{Interface construction}
Given an adsorbate+surface structure, the solid-liquid interface is constructed by sampling a random solvent and ion combination. Solvents are sampled from a list of eight commonly used solvents (e.g. polar/nonpolar, protic/aprotic). Similarly, ions are sampled from nine cations and anions of varying charges and sizes. The full list of solvents and ions is illustrated in Supplementary Figure \ref{fig:oc25_stats}. The surface charge density distribution of the metal interfaces sampled in the dataset is shown in Supplementary Figure \ref{fig:oc25_stats}, with values ranging from $\sim$ -80 µC/cm$^2$ to $\sim$ 60 µC/cm$^2$, corresponding to cathodic (reducing) and anodic (oxidizing) conditions, respectively. As can be seen in the ion and surface charge density distributions, a majority of the interfaces have zero surface charge density which corresponds to the condition of the potential of zero charge (PZC).

Given the importance and frequent use of water as a solvent in electrocatalytic applications, we biased our sampling towards water, with all other solvents uniformly weighted. An ion is only sampled $\sim$50\% of the time. A solvent depth is then sampled between 5 and 10\angs, with more weighting on $\leq$6\angs~to limit excessive computational cost. Given the solvent depth and area of the surface, N solvent molecules are selected, where N is the number of molecules necessary to approximately satisfy the density of the solvent. The resulting solvent+ion box is then randomly packed with \texttt{Packmol}~\cite{martinez2009packmol} and placed on top of the adsorbate+surface configuration. 

To better capture meaningful interactions, for a subset of the dataset, we pre-optimize initial geometries with existing OC20-trained models. EquiformerV2-31M~\cite{liao2023equiformerv2} and UMA-S-1~\cite{wood2025family} were used to relax geometries using loose convergence criterion of maximum per-atom force of 0.5 eV/\angs~or 50 steps, whichever comes first.

\subsubsection{Ab initio calculations}
Configurations are then evaluated with DFT in one of two ways: relaxations or \textit{ab initio} molecular dynamics (AIMD). Structures sampled for relaxations are optimized for only 5 ionic steps. Similarly, short-time scale (10-50 steps) AIMD simulations are performed at constant temperature and volume (NVT) at a temperature of 1000K. We limit simulations to short-time scales to maximize diversity in the dataset. 

All DFT calculations were performed with the Vienna Ab Initio Simulation Package (VASP)~\cite{Kresse1994, Kresse1996, Kresse1996a, kresse1999ultrasoft} v6.3.2. Similar to other large-scale dataset efforts (OC20/OC22), a broad set of settings was selected to balance accuracy and computational costs. Calculations were performed with a revised Perdew-Burke-Ernzerhof (RPBE) functional \cite{hammer1999improved}, supplemented with the D3 correction with zero damping to account for the non-local van der Waals dispersion interactions~\cite{grimme2010consistent}, plane wave cutoff energy of 400 eV, and a dipole correction in the z-direction \cite{bengtsson1999dipole}. The k-point mesh was constructed as a function of the cell parameters, similar to OC20, using a reciprocal density of 40.  We utilized the non spin-polarized RPBE functional for two reasons: first, the vast majority of the surfaces sampled in this dataset are not magnetic, meaning enabling spin polarization would only add computational expense to dataset generation. Second, for the systems in our database that are magnetic and thus could benefit from spin polarization, ensuring the correct magnetic behavior (ferromagnetism vs antiferromagnetism, etc.) is not trivial, and thus the incremental benefit was deemed not to be worth the additional computational cost. The full set of VASP parameters can be found at \href{https://github.com/facebookresearch/fairchem/blob/dba2f17b46273533eccd7768a8a0fd028878fe86/src/fairchem/data/oc/utils/vasp_flags.py#L47-L103}{https://github.com/facebookresearch/fairchem/}.

\subsubsection{Force convergence and consistency}

The consistency between energy and force (e.g., $F=- \frac{dE}{dx}$) labels in DFT is critical to building reliable datasets for MLIP training. Given a level of theory, a static calculation in DFT codes like VASP is considered complete when the electronic self-consistency loop is converged. Convergence is often defined based on a break condition in total energy (e.g., total energy change between two electronic steps < $\sigma$, ``EDIFF'' in VASP). For OC25, the electronic termination criterion for the training data was set to $10^{-4}$ eV, balancing accuracy and computational cost, similar to previous works \cite{oc20,oc22,omat24}. 

The net force on any system is expected to be identically zero (i.e., zero acceleration) in the absence of any external fields. However, if the electronic structure calculations are not fully converged, non-zero net forces may be observed. DFT codes often implement routines to correct for these spurious forces, denoted as `force drift.' For MD calculations in VASP, this drift is calculated and removed from the system during the integration step. However, only the magnitude of the force drift is saved as an output, not the corrected force actually used within the simulation\footnote{https://www.vasp.at/wiki/index.php/Category:Forces}. For the validation and test datasets, all calculations were conducted as single points (i.e., non-MD calculations, IBRION$\neq0$) where the spurious drift was removed using a tighter EDIFF in VASP ($10^{-6}$~eV). This allowed us to directly test the impact of force convergence on final model performance, both for assessing the value of OC25 and for informing whether future dataset efforts require tighter convergence thresholds.

To determine which samples to include in the final training dataset, we rigorously tested the impact of the net-force drift correction and the ``EDIFF'' threshold for force convergence on a subset of the dataset. In Supplementary Figure \ref{fig:drift}, we show that the magnitude of the drift correction is strongly correlated with the error between the forces determined using more tightly converged calculations (EDIFF=10$^{-6}$~eV) and the forces determined using training data settings (EDIFF=10$^{-4}$~eV), and that structures with total drift greater than 10 eV/\angs~have much larger errors. Energy errors are largely unaffected, with total energy errors as low as 1.5 meV. To ensure reliable forces, we chose a conservative threshold of 1 eV/\angs~ force drift as a quality criterion. The final training dataset was filtered to include only calculations with drifts smaller than this value. 

We also investigated the ability of models trained on the unfiltered training data to predict the forces of more tightly converged (EDIFF=10$^{-6}$~eV) calculations in the validation/test datasets. Surprisingly, we found that models trained on the less tightly converged data were able to predict the forces of the more tightly converged set (see Supplementary Figure \ref{fig:parity_nodrift}). This suggests that the models are to some degree robust to noise in the training data, and is somewhat at odds with the conventional wisdom that any noise in the training data will lead to reduced model performance.

\subsubsection{Training splits}
The OC25 dataset is divided into training, validation, and test splits to ensure consistent evaluations by the community. Splits are created based on unique bulk-solvent combinations. Of the $\sim$260,000 unique pairings, $\sim$2.5\% each are held out for validation and test. For each data point, DFT total energies and per-atom forces are provided. To test generalizability beyond our defined splits, we generate several explicit out-of-distribution (OOD) splits. Dataset splits are summarized in Supplementary Table \ref{tab:dataset}.

\textbf{Solvents}.
To assess model performance beyond the solvents used for OC25, a few additional solvents were sampled. These include ethylene carbonate, acetonitrile, ethanol, and dichloromethane. Bulk materials, adsorbates, and ions used in these configurations remain in-distribution.

\textbf{Ions}.
Ions beyond the ones used in OC25 were also sampled to evaluate generalizability on. These include \ce{Cl-}, \ce{PO4^3-}, \ce{Mg^2+}, and \ce{NO3-}. Bulk materials, adsorbates, and solvents used in these configurations remain in distribution.

\textbf{Solvents + Ions (Both)}.
Structures with both OOD solvents and ions compose this split. Only bulk materials and adsorbates used in these configurations remain in distribution.

\textbf{Interfacial solvation}.
Adsorbate solvation energy is a commonly used property for studying how solvents affect the interactions between adsorbed reaction intermediates and molecules, ions, or complexes in solution~\cite{heenen2020solvation}. In the context of catalysis, this often corresponds to the energy difference between the adsorption energy in a solvated environment and in vacuum: 

\begin{equation}
    \Delta E_{solv} = \Delta E_{ads}^{solv} - \Delta E_{ads}^{vac}
\end{equation}

As a proxy for this metric, we perform single-point DFT evaluations of solvated configurations, and delete the respective regions (e.g. solvent, adsorbate, solvent+adsorbate) to generate the reference configurations. A pseudo-solvation energy, $\tilde{\Delta E_{solv}}$, is calculated based on these static snapshots. We omit any relaxations or molecular dynamics steps that would typically follow, in order to simplify the task and ensure deterministic MLIP evaluation.

\subsection{Baseline models}
We evaluate OC25 using a set of baseline models that represent state-of-the-art models for catalysis. Baseline models include UMA~\cite{wood2025family} and eSEN~\cite{eSEN}, graph neural networks (GNNs) that operate on graphs where atoms are nodes and edges are the interactions between them. We evaluate models of different sizes as well as energy-conserving and direct-force models. We also evaluate the performance of fine-tuning from the latest energy-conserving UMA model (oc20 head). Generally, models demonstrate competitive performance, with energy and force errors as low as 0.10 eV and 0.015 eV/Å for eSEN-S-cons.

Baseline results for all splits are evaluated using mean absolute errors (MAE) of energy and forces as primary metrics. Results across the different test sets are provided in Table \ref{tab:test-results}. All evaluation sets are calculated using tighter DFT convergence criteria (EDIFF=$10^{-6}$~eV) to ensure more accurate force labels. Model hyperparameters and training details can be found in Supplementary Table \ref{tab:hyperparams}.

\begin{table}[t!]
\caption{
Baseline results across the different \textbf{test} splits for different graph neural network models defined in the text. Energy and force mean absolute errors (MAE) are reported in units of eV and eV/\angs. Validation results are provided in the Appendix.
}
\label{tab:test-results}
\resizebox{\textwidth}{!}{%
\begin{tabular}{@{}llrccccccccc@{}}
\toprule
 & &  & \multicolumn{2}{c}{Test} & \multicolumn{2}{c}{OOD Solvent} & \multicolumn{2}{c}{OOD Ion} & \multicolumn{2}{c}{OOD Both} & Solvation \\ \midrule
Dataset & Model & \# of params & Energy & Forces & Energy & Forces & Energy & Forces & Energy & Forces & Energy \\ \midrule
\multirow{3}{*}{OC25} 
& eSEN-S-d. & 6.3M & 0.138 & 0.020 & 0.351 & 0.047 & 0.216 & 0.035 & 0.389 & 0.052 & 0.060 \\
& eSEN-S-cons. & 6.3M & 0.105 & 0.015 & \textbf{0.175} & 0.035 & 0.143 & 0.026 & \textbf{0.186} & 0.038 & 0.045 \\
& eSEN-M-d. & 50.7M & \textbf{0.060} & \textbf{0.009} & 0.238 & \textbf{0.023} & \textbf{0.122} & \textbf{0.018} & 0.264 & \textbf{0.026} & \textbf{0.040} \\ \midrule
UMA & UMA-S-1.1 & 146.6M & - & 0.064 & - & 0.101 & - & 0.090 & - & 0.108 & 0.169 \\
UMA$\rightarrow$OC25 & UMA-S-ft & 146.6M & 0.091 & 0.014 & 0.201 & 0.036 & 0.148 & 0.027 & 0.225 & 0.039 & 0.136 \\ \bottomrule
\end{tabular}%
}
\end{table}

\subsection{Literature simulations}

The interfacial structure benchmarks follow the simulation protocol of Dom\'inguez-Flores et al.~\cite{dominguez2024metal}. Five-layer close-packed $6\times6$ supercells of Rh(111), Pd(111), Pt(111), Ag(111), Au(111), and Ru(0001) were solvated with 144 explicit water molecules and simulated in the NVT ensemble at 298~K using a Langevin thermostat with a 1~fs timestep for approximately 100~ps.

Metadynamics simulations of CO dimerization were performed on a $3\times4$ Cu(100)/water interface, closely matching the settings of Kristoffersen and Chan~\cite{kristoffersen2021towards} despite differences in the underlying engine code (VASP versus ASE/PLUMED). The C--C distance between the two adsorbed CO molecules was used as the collective variable. The CV was biased by the periodic addition of repulsive Gaussian potentials every 10 steps (deposition time of 10~fs), each initialized with a height of 0.001~eV and a width of $\sigma = 0.06$~\AA{}. A timestep of 1~fs was used. To probe double-layer charging, two Cs$^+$ ions were introduced at the Cu(100)/water interface, imposing an effective negative surface charge density of $-41.8~\mu$C/cm$^2$, similar to Kristoffersen and Chan~\cite{kristoffersen2021towards}. Work functions for the initial, transition, and final states were computed from $\sim$50 random single-point DFT calculations sampled from each state along the eSEN-OC25 MD trajectories.

\subsection{Enhanced sampling}
The free energy profiles for CO dimerization are calculated using the On-the-fly Probability Enhanced Sampling (OPES) method~\cite{opes} implemented in the open-source, community-developed PLUMED library~\cite{plumed}, interfaced with ASE's molecular dynamics engine~\cite{ase}. The collective variable (CV) is defined as the distance between the two carbon atoms of the adsorbed CO molecules, with an upper wall potential applied at 6\AA{} ($\kappa=100$~eV/\angs$^{2}$). Adaptive kernels (\texttt{SIGMA=ADAPTIVE} in PLUMED) were used so that the Gaussian widths are tuned on the fly from the running variance of the CV, removing the need to fix a single \texttt{SIGMA} a priori. A detailed description of the frequency and barrier heights of the adaptive Gaussian kernels being deposited are provided in Table \ref{tab:opes_sweep}. The equations of motion were propagated using a Langevin thermostat at 300 K with a 0.5 fs timestep. Each simulation was run for 15 million steps ($\sim$7.5 ns total).

\begin{table}[!htbp]
\centering
\caption{OPES hyperparameter sweep grid per surface and ion identity. Each
listed (barrier, pace) combination corresponds to a separate 7.5 ns
trajectory. A broader (barrier, pace) grid was used for Cs and Li on Cu(310) to ensure adequate sampling of the CV space along the stepped facet.}
\label{tab:opes_sweep}
\begin{tabular}{l l c c}
\toprule
Surface & Ion & BARRIER (eV) & PACE\\
\midrule
8$\times$8 Cu(100)      & Cs, Li, K & $2, 5$  & $100, 500, 1000$       \\
\addlinespace
6$\times$8 Cu(100)      & Cs, Li, K & $2, 5$   & $100, 500, 1000$       \\
\addlinespace
\multirow{2}{*}{Cu(310)} & Cs, Li & $0.1, 0.3, 0.5, 2, 5, 8$ & $10, 100, 500, 1000$ \\
   & K   & $2, 5$    & $100, 500, 1000$       \\
\bottomrule
\end{tabular}
\end{table}

Free energy surfaces (FES) along the CV were obtained by reweighting on-the-fly biased
trajectories generated with OPES. For each production trajectory, 10\% of the simulation was discarded as equilibration time. Statistical uncertainties were estimated by partitioning each trajectory into $N_b=5$ contiguous blocks, computing the reweighted FES independently in each block, and propagating the variance of the block-resolved probabilities to the free energy following the standard OPES error-estimation procedure. To assess convergence, we additionally swept $N_b=10$ and the initial equilibration discard fraction (5\%, 10\%, 20\% of the trajectory), finding minimal differences across the set.

\subsection{Implicit model}
For the CO dimerization energetics that utilized a polarizable continuum model to represent the electrolyte (cf. Figure \ref{fig:dimerization_surf_charge}(c)), we used the VASPsol++ implementation of Plaisance and coworkers~\cite{islam2023implicit}. Specifically, we used the linear solvation model, which determines countercharge placement by solving the linearized Poisson-Boltzmann equation. A 4$\times$4 supercell of Cu (100) was utilized, using the same functional (RPBE-D3), k-point density (for this supercell size, 3$\times$3$\times$1), and bulk lattice constant as the training data used for developing OC25. Transition states were determined using the dimer method as implemented by Henkelman and coworkers~\cite{henkelman1999dimer}. Entropic contributions to both geometric minima and first-order saddle points were determined via the harmonic oscillator approximation, which neglects any translational and rotational degrees of freedom, and simultaneously approximates all vibrational degrees of freedom as fully harmonic. Additionally, the surface charge density (c.f. the x-axis of Figure \ref{fig:dimerization_surf_charge}(c)) was divided by two, since in the continuum solvation model, both sides of the surface slab are charged. 

\subsection{DFT validation of MLIP configurations}

To assess the reliability of the eSEN-OC25 model for the large-cell CO dimerization simulations, which represent an out-of-distribution regime (larger systems, multiple ions, more complex interactions than training data), we compute DFT single-points on ML-generated configurations. For each simulation, $\sim$100 configurations are sampled from the initial, transition, and final states to evaluate energy and force errors. Barrier and reaction energy errors are computed from the mean energy of each state, with mean absolute errors (MAE) reported for forces across all configurations. Additionally, the evaluated total energy errors exhibit a systematic offset that scales linearly with the cation count. This offset is constant within each system configuration and cancels exactly in all energy differences. Fitting a per-ion correction remedies this and is shown in Supplementary Table~\ref{si:tab:ml_vs_dft}.

\clearpage
\newpage
\bibliographystyle{assets/plainnat}
\bibliography{paper}

@article{Monteiro2021,
  author  = {Monteiro, Mariana C. O. and Dattila, Federico and Hagedoorn, Bellis and Garc{\'i}a-Muelas, Rodrigo and L{\'o}pez, N{\'u}ria and Koper, Marc T. M.},
  title   = {Absence of {CO$_2$} Electroreduction on Copper, Gold and Silver Electrodes without Metal Cations in Solution},
  journal = {Nature Catalysis},
  volume  = {4},
  number  = {8},
  pages   = {654--662},
  year    = {2021},
  doi     = {10.1038/s41929-021-00655-5}
}

@article{Xu2026,
  author  = {Xu, Yifei and Zhao, Kaiyue and Chang, Xiaoxia and Xu, Bingjun},
  title   = {Emerging Roles of Cations in Electrocatalytic Reduction of {CO$_2$} and {CO}},
  journal = {Nature Energy},
  volume  = {11},
  pages   = {387--399},
  year    = {2026},
  doi     = {10.1038/s41560-026-01973-3}
}

@article{schiffer2017electrification,
  title={Electrification and decarbonization of the chemical industry},
  author={Schiffer, Zachary J and Manthiram, Karthish},
  journal={Joule},
  volume={1},
  number={1},
  pages={10--14},
  year={2017},
  publisher={Elsevier}
}

@article{mallapragada2023decarbonization,
  title={Decarbonization of the chemical industry through electrification: Barriers and opportunities},
  author={Mallapragada, Dharik S and Dvorkin, Yury and Modestino, Miguel A and Esposito, Daniel V and Smith, Wilson A and Hodge, Bri-Mathias and Harold, Michael P and Donnelly, Vincent M and Nuz, Alice and Bloomquist, Casey and others},
  journal={Joule},
  volume={7},
  number={1},
  pages={23--41},
  year={2023},
  publisher={Elsevier}
}

@article{barecka2023towards,
  title={Towards an accelerated decarbonization of the chemical industry by electrolysis},
  author={Barecka, Magda H and Ager, Joel W},
  journal={Energy Advances},
  volume={2},
  number={2},
  pages={268--279},
  year={2023},
  publisher={Royal Society of Chemistry}
}

@article{cabana2022ngene,
  title={NGenE 2022: Electrochemistry for decarbonization},
  author={Cabana, Jordi and Alaan, Thomas and Crabtree, George W and Huang, Po-Wei and Jain, Akash and Murphy, Megan and N’Diaye, Jeanne and Ojha, Kasinath and Agbeworvi, George and Bergstrom, Helen and others},
  journal={ACS Energy Letters},
  volume={8},
  number={1},
  pages={740--747},
  year={2022},
  publisher={ACS Publications}
}

@article{xia2022emerging,
  title={Emerging electrochemical processes to decarbonize the chemical industry},
  author={Xia, Rong and Overa, Sean and Jiao, Feng},
  journal={JACS Au},
  volume={2},
  number={5},
  pages={1054--1070},
  year={2022},
  publisher={ACS Publications}
}

@article{miao2023electrified,
  title={Electrified cement production via anion-mediated electrochemical calcium extraction},
  author={Miao, Rui Kai and Wang, Ning and Hung, Sung-Fu and Huang, Wen-Yang and Zhang, Jinqiang and Zhao, Yong and Ou, Pengfei and Wang, Sasa and Edwards, Jonathan P and Tian, Cong and others},
  journal={ACS Energy Letters},
  volume={8},
  number={11},
  pages={4694--4701},
  year={2023},
  publisher={ACS Publications}
}

@article{zhang2025electrolytic,
  title={Electrolytic cement clinker precursor production sustained through orthogonalization of ion vectors},
  author={Zhang, Zishuai and Williams, Aubry SR and Ren, Shaoxuan and Mowbray, Benjamin AW and Parkyn, Colin TE and Kim, Yongwook and Ji, Tengxiao and Berlinguette, Curtis P},
  journal={Energy \& Environmental Science},
  year={2025},
  volume={18},
  number={5},
  pages={2395--2404},
  publisher={Royal Society of Chemistry}
}

@article{chung2024direct,
  title={Direct propylene epoxidation via water activation over Pd-Pt electrocatalysts},
  author={Chung, Minju and Maalouf, Joseph H and Adams, Jason S and Jiang, Chenyu and Rom{\'a}n-Leshkov, Yuriy and Manthiram, Karthish},
  journal={Science},
  volume={383},
  number={6678},
  pages={49--55},
  year={2024},
  publisher={American Association for the Advancement of Science}
}

@article{zhang2023advances,
  title={Advances in electrochemical oxidation of olefins to epoxides},
  author={Zhang, Peng and Wang, Tuo and Gong, Jinlong},
  journal={CCS Chemistry},
  volume={5},
  number={5},
  pages={1028--1042},
  year={2023},
  publisher={Chinese Chemical Society Zhongguancun, Haidian, Beijing 100190, China}
}

@article{wang2023electrochemical,
  title={Electrochemical Epoxidation of Propylene to Propylene Oxide via Halogen-Mediated Systems},
  author={Wang, Jiangjiang and Wu, Gangfeng and Feng, Guanghui and Li, Guihua and Wei, Yiheng and Li, Shoujie and Mao, Jianing and Liu, Xiaohu and Chen, Aohui and Song, Yanfang and others},
  journal={ACS Omega},
  volume={8},
  number={49},
  pages={46569--46576},
  year={2023},
  publisher={ACS Publications}
}

@incollection{gouda2024green,
  title={Green Hydrogen Production: From Lab Scale to Pilot Scale Photocatalysis},
  author={Gouda, Ashrumochan and Sharma, Devendra and Kumar, Ashish and Krishnan, Venkata},
  booktitle={Towards Sustainable and Green Hydrogen Production by Photocatalysis: Scalability Opportunities and Challenges (Volume 1)},
  pages={185--210},
  year={2024},
  publisher={ACS Publications}
}

@article{wu2022energy,
  title={Energy decarbonization via green H$_2$ or NH$_3$?},
  author={Wu, Simson and Salmon, Nicholas and Li, Molly Meng-Jung and Ba{\~n}ares-Alc{\'a}ntara, Ren{\'e} and Tsang, Shik Chi Edman},
  journal={ACS Energy Letters},
  volume={7},
  number={3},
  pages={1021--1033},
  year={2022},
  publisher={ACS Publications}
}

@article{tao2022engineering,
  title={Engineering challenges in green hydrogen production systems},
  author={Tao, Meng and Azzolini, Joseph A and Stechel, Ellen B and Ayers, Katherine E and Valdez, Thomas I},
  journal={Journal of The Electrochemical Society},
  volume={169},
  number={5},
  pages={054503},
  year={2022},
  publisher={IOP Publishing}
}

@article{olusegun2024understanding,
  title={Understanding activity trends in electrochemical dinitrogen oxidation over transition metal oxides},
  author={Olusegun, Samuel A and Qi, Yancun and Kani, Nishithan C and Singh, Meenesh R and Gauthier, Joseph A},
  journal={ACS Catalysis},
  volume={14},
  number={22},
  pages={16885--16896},
  year={2024},
  publisher={ACS Publications}
}

@article{iriawan2021methods,
  title={Methods for nitrogen activation by reduction and oxidation},
  author={Iriawan, Haldrian and Andersen, Suzanne Z and Zhang, Xilun and Comer, Benjamin M and Barrio, Jes{\'u}s and Chen, Ping and Medford, Andrew J and Stephens, Ifan EL and Chorkendorff, Ib and Shao-Horn, Yang},
  journal={Nature Reviews Methods Primers},
  volume={1},
  number={1},
  pages={56},
  year={2021},
  publisher={Nature Publishing Group UK London}
}

@article{lazouski2019understanding,
  title={Understanding continuous lithium-mediated electrochemical nitrogen reduction},
  author={Lazouski, Nikifar and Schiffer, Zachary J and Williams, Kindle and Manthiram, Karthish},
  journal={Joule},
  volume={3},
  number={4},
  pages={1127--1139},
  year={2019},
  publisher={Elsevier}
}

@article{fu2024calcium,
  title={Calcium-mediated nitrogen reduction for electrochemical ammonia synthesis},
  author={Fu, Xianbiao and Niemann, Valerie A and Zhou, Yuanyuan and Li, Shaofeng and Zhang, Ke and Pedersen, Jakob B and Saccoccio, Mattia and Andersen, Suzanne Z and Enemark-Rasmussen, Kasper and Benedek, Peter and others},
  journal={Nature Materials},
  volume={23},
  number={1},
  pages={101--107},
  year={2024},
  publisher={Nature Publishing Group UK London}
}

@article{goyal2024metal,
  title={Metal Nitride as a Mediator for the Electrochemical Synthesis of NH$_3$},
  author={Goyal, Ishita and Kani, Nishithan C and Olusegun, Samuel A and Chinnabattigalla, Sreenivasulu and Bhawnani, Rajan R and Glusac, Ksenija D and Singh, Aayush R and Gauthier, Joseph A and Singh, Meenesh R},
  journal={ACS Energy Letters},
  volume={9},
  number={8},
  pages={4188--4195},
  year={2024},
  publisher={ACS Publications}
}

@article{wang2024fast,
  title={``Fast-charging'' anode materials for lithium-ion batteries from perspective of ion diffusion in crystal structure},
  author={Wang, Rui and Wang, Lu and Liu, Rui and Li, Xiangye and Wu, Youzhi and Ran, Fen},
  journal={ACS Nano},
  volume={18},
  number={4},
  pages={2611--2648},
  year={2024},
  publisher={ACS Publications}
}

@article{du2024side,
  title={Side reactions/changes in lithium-ion batteries: mechanisms and strategies for creating safer and better batteries},
  author={Du, Hao and Wang, Yadong and Kang, Yuqiong and Zhao, Yun and Tian, Yao and Wang, Xianshu and Tan, Yihong and Liang, Zheng and Wozny, John and Li, Tao and others},
  journal={Advanced Materials},
  volume={36},
  number={29},
  pages={2401482},
  year={2024},
  publisher={Wiley Online Library}
}

@article{liu2024revealing,
  title={Revealing the degradation patterns of lithium-ion batteries from impedance spectroscopy using variational auto-encoders},
  author={Liu, Yanshuo and Li, Qiang and Wang, Kai},
  journal={Energy Storage Materials},
  volume={69},
  pages={103394},
  year={2024},
  publisher={Elsevier}
}

@article{bai2024low,
  title={Low-Temperature Sodium-Ion Batteries: Challenges and Progress},
  author={Bai, Zhongchao and Yao, Qian and Wang, Mingyue and Meng, Weijia and Dou, Shixue and Liu, Hua kun and Wang, Nana},
  journal={Advanced Energy Materials},
  volume={14},
  number={17},
  pages={2303788},
  year={2024},
  publisher={Wiley Online Library}
}

@article{zhang2024emerging,
  title={Emerging chemistry for wide-temperature sodium-ion batteries},
  author={Zhang, Fang and He, Bijiao and Xin, Yan and Zhu, Tiancheng and Zhang, Yuning and Wang, Shuwei and Li, Weiyi and Yang, Yang and Tian, Huajun},
  journal={Chemical Reviews},
  volume={124},
  number={8},
  pages={4778--4821},
  year={2024},
  publisher={ACS Publications}
}

@article{cai2024challenges,
  title={Challenges and industrial perspectives on the development of sodium ion batteries},
  author={Cai, Xiaosheng and Yue, Yingying and Yi, Zheng and Liu, Junfei and Sheng, Yangping and Lu, Yuhao},
  journal={Nano Energy},
  volume={129},
  pages={110052},
  year={2024},
  publisher={Elsevier}
}

@article{li2024recent,
  title={Recent advances in molecularly imprinted polymer-based electrochemical sensors},
  author={Li, Yixuan and Luo, Liuxiong and Kong, Yingqi and Li, Yujia and Wang, Quansheng and Wang, Mingqing and Li, Ying and Davenport, Andrew and Li, Bing},
  journal={Biosensors and Bioelectronics},
  volume={249},
  pages={116018},
  year={2024},
  publisher={Elsevier}
}

@article{hao2025recent,
  title={Recent advancements in electrochemical sensors based on MOFs and their derivatives},
  author={Hao, Xi and Song, Weihua and Wang, Yinghui and Qin, Jieling and Jiang, Zhenqi},
  journal={Small},
  volume={21},
  number={4},
  pages={2408624},
  year={2025},
  publisher={Wiley Online Library}
}

@article{singh2024review,
  title={A review on recent trends and future developments in electrochemical sensing},
  author={Singh, Rimmy and Gupta, Ruchi and Bansal, Deepak and Bhateria, Rachna and Sharma, Mona},
  journal={ACS Omega},
  volume={9},
  number={7},
  pages={7336--7356},
  year={2024},
  publisher={ACS Publications}
}

@article{heenen2020solvation,
  title={Solvation at metal/water interfaces: An ab initio molecular dynamics benchmark of common computational approaches},
  author={Heenen, Hendrik H and Gauthier, Joseph A and Kristoffersen, Henrik H and Ludwig, Thomas and Chan, Karen},
  journal={The Journal of Chemical Physics},
  volume={152},
  number={14},
  year={2020},
  publisher={AIP Publishing}
}

@article{govindarajan2025intricacies,
author = {Govindarajan, Nitish and Kastlunger, Georg and Gauthier, Joseph A. and Cheng, Jun and Filot, Ivo and Hagopian, Arthur and Hansen, Heine Anton and Huang, Jun and Kowalski, Piotr M. and Liu, Jinwen and Lombardi, Juan M. and Maraschin, Mikael and Peterson, Andrew and Pillai, Hemanth S. and Prats, Hector and Price, Conor J. and van Roij, Ren{\'e} and Rossmeisl, Jan and Seemakurthi, Ranga Rohit and Shin, Seung-Jae and Smith, Audrey and Zhu, Jia-Xin and Doblhoff-Dier, Katharina},
title = {The Intricacies of Computational Electrochemistry},
journal = {ACS Energy Letters},
volume = {10},
number = {9},
pages = {4277-4288},
year = {2025},
publisher={ACS Publications}
}

@article{oc20,
  title={Open catalyst 2020 (OC20) dataset and community challenges},
  author={Chanussot, Lowik and Das, Abhishek and Goyal, Siddharth and Lavril, Thibaut and Shuaibi, Muhammed and Riviere, Morgane and Tran, Kevin and Heras-Domingo, Javier and Ho, Caleb and Hu, Weihua and others},
  journal={ACS Catalysis},
  volume={11},
  number={10},
  pages={6059--6072},
  year={2021},
  publisher={ACS Publications}
}

@article{oc22,
  title={The Open Catalyst 2022 (OC22) dataset and challenges for oxide electrocatalysts},
  author={Tran, Richard and Lan, Janice and Shuaibi, Muhammed and Wood, Brandon M and Goyal, Siddharth and Das, Abhishek and Heras-Domingo, Javier and Kolluru, Adeesh and Rizvi, Ammar and Shoghi, Nima and others},
  journal={ACS Catalysis},
  volume={13},
  number={5},
  pages={3066--3084},
  year={2023},
  publisher={ACS Publications}
}

@article{liao2023equiformerv2,
  title={Equiformerv2: Improved equivariant transformer for scaling to higher-degree representations},
  author={Liao, Yi-Lun and Wood, Brandon and Das, Abhishek and Smidt, Tess},
  journal={arXiv preprint arXiv:2306.12059},
  year={2023}
}

@article{gasteiger2022gemnet,
  title={GemNet-OC: developing graph neural networks for large and diverse molecular simulation datasets},
  author={Gasteiger, Johannes and Shuaibi, Muhammed and Sriram, Anuroop and G{\"u}nnemann, Stephan and Ulissi, Zachary and Zitnick, C Lawrence and Das, Abhishek},
  journal={arXiv preprint arXiv:2204.02782},
  year={2022}
}

@incollection{jain2020materials,
  title={The materials project: Accelerating materials design through theory-driven data and tools},
  author={Jain, Anubhav and Montoya, Joseph and Dwaraknath, Shyam and Zimmermann, Nils ER and Dagdelen, John and Horton, Matthew and Huck, Patrick and Winston, Donny and Cholia, Shreyas and Ong, Shyue Ping and others},
  booktitle={Handbook of Materials Modeling: Methods: Theory and Modeling},
  pages={1751--1784},
  year={2020},
  publisher={Springer}
}

@article{omat24,
  title={Open materials 2024 (omat24) inorganic materials dataset and models},
  author={Barroso-Luque, Luis and Shuaibi, Muhammed and Fu, Xiang and Wood, Brandon M and Dzamba, Misko and Gao, Meng and Rizvi, Ammar and Zitnick, C Lawrence and Ulissi, Zachary W},
  journal={arXiv preprint arXiv:2410.12771},
  year={2024}
}

@article{ringe2021implicit,
author = {Ringe, Stefan and H{\"o}rmann, Nicolas G. and Oberhofer, Harald and Reuter, Karsten},
title = {Implicit Solvation Methods for Catalysis at Electrified Interfaces},
journal = {Chemical Reviews},
volume = {122},
number = {12},
pages = {10777-10820},
year = {2022},
}

@article{wander2025cattsunami,
  title={CatTSunami: Accelerating transition state energy calculations with pretrained graph neural networks},
  author={Wander, Brook and Shuaibi, Muhammed and Kitchin, John R and Ulissi, Zachary W and Zitnick, C Lawrence},
  journal={ACS Catalysis},
  volume={15},
  number={7},
  pages={5283--5294},
  year={2025},
  publisher={ACS Publications}
}

@article{abed2024open,
  title={Open catalyst experiments 2024 (OCx24): Bridging experiments and computational models},
  author={Abed, Jehad and Kim, Jiheon and Shuaibi, Muhammed and Wander, Brook and Duijf, Boris and Mahesh, Suhas and Lee, Hyeonseok and Gharakhanyan, Vahe and Hoogland, Sjoerd and Irtem, Erdem and others},
  journal={arXiv preprint arXiv:2411.11783},
  year={2024}
}

@article{wood2025family,
  title={UMA: A Family of Universal Models for Atoms},
  author={Wood, Brandon M and Dzamba, Misko and Fu, Xiang and Gao, Meng and Shuaibi, Muhammed and Barroso-Luque, Luis and Abdelmaqsoud, Kareem and Gharakhanyan, Vahe and Kitchin, John R and Levine, Daniel S and others},
  journal={arXiv preprint arXiv:2506.23971},
  year={2025}
}

@article{hammer1999improved,
  title={Improved adsorption energetics within density-functional theory using revised Perdew-Burke-Ernzerhof functionals},
  author={Hammer, Bj{\o}rk and Hansen, Lars Bruno and N{\o}rskov, Jens Kehlet},
  journal={Physical Review B},
  volume={59},
  number={11},
  pages={7413},
  year={1999},
  publisher={APS}
}

@article{bengtsson1999dipole,
  title={Dipole correction for surface supercell calculations},
  author={Bengtsson, Lennart},
  journal={Physical Review B},
  volume={59},
  number={19},
  pages={12301},
  year={1999},
  publisher={APS}
}

@article{eSEN,
  title={Learning smooth and expressive interatomic potentials for physical property prediction},
  author={Fu, Xiang and Wood, Brandon M and Barroso-Luque, Luis and Levine, Daniel S and Gao, Meng and Dzamba, Misko and Zitnick, C Lawrence},
  journal={arXiv preprint arXiv:2502.12147},
  year={2025}
}

@article{omol,
  title={The open molecules 2025 (omol25) dataset, evaluations, and models},
  author={Levine, Daniel S and Shuaibi, Muhammed and Spotte-Smith, Evan Walter Clark and Taylor, Michael G and Hasyim, Muhammad R and Michel, Kyle and Batatia, Ilyes and Cs{\'a}nyi, G{\'a}bor and Dzamba, Misko and Eastman, Peter and others},
  journal={arXiv preprint arXiv:2505.08762},
  year={2025}
}

@article{adamw,
  title={Decoupled weight decay regularization},
  author={Loshchilov, Ilya and Hutter, Frank},
  journal={arXiv preprint arXiv:1711.05101},
  year={2017}
}

@article{qm9,
  title={Quantum chemistry structures and properties of 134 kilo molecules},
  author={Ramakrishnan, Raghunathan and Dral, Pavlo O and Rupp, Matthias and Von Lilienfeld, O Anatole},
  journal={Scientific Data},
  volume={1},
  number={1},
  pages={1--7},
  year={2014},
  publisher={Nature Publishing Group}
}

@article{eastman2023spice,
  title={Spice, a dataset of drug-like molecules and peptides for training machine learning potentials},
  author={Eastman, Peter and Behara, Pavan Kumar and Dotson, David L and Galvelis, Raimondas and Herr, John E and Horton, Josh T and Mao, Yuezhi and Chodera, John D and Pritchard, Benjamin P and Wang, Yuanqing and others},
  journal={Scientific Data},
  volume={10},
  number={1},
  pages={11},
  year={2023},
  publisher={Nature Publishing Group UK London}
}

@article{deng2023chgnet,
  title={CHGNet as a pretrained universal neural network potential for charge-informed atomistic modelling},
  author={Deng, Bowen and Zhong, Peichen and Jun, KyuJung and Riebesell, Janosh and Han, Kevin and Bartel, Christopher J and Ceder, Gerbrand},
  journal={Nature Machine Intelligence},
  volume={5},
  number={9},
  pages={1031--1041},
  year={2023},
  publisher={Nature Publishing Group UK London}
}

@article{martinez2009packmol,
  title={PACKMOL: A package for building initial configurations for molecular dynamics simulations},
  author={Mart{\'\i}nez, Leandro and Andrade, Ricardo and Birgin, Ernesto G and Mart{\'\i}nez, Jos{\'e} Mario},
  journal={Journal of Computational Chemistry},
  volume={30},
  number={13},
  pages={2157--2164},
  year={2009},
  publisher={Wiley Online Library}
}

@article{lan2023adsorbml,
  title={AdsorbML: A leap in efficiency for adsorption energy calculations using generalizable machine learning potentials},
  author={Lan, Janice and Palizhati, Aini and Shuaibi, Muhammed and Wood, Brandon M and Wander, Brook and Das, Abhishek and Uyttendaele, Matt and Zitnick, C Lawrence and Ulissi, Zachary W},
  journal={npj Computational Materials},
  volume={9},
  number={1},
  pages={172},
  year={2023},
  publisher={Nature Publishing Group UK London}
}

@article{levell2024emerging,
author = {Levell, Zachary and Le, Jiabo and Yu, Saerom and Wang, Ruoyu and Ethirajan, Sudheesh and Rana, Rachita and Kulkarni, Ambarish and Resasco, Joaquin and Lu, Deyu and Cheng, Jun and Liu, Yuanyue},
title = {Emerging Atomistic Modeling Methods for Heterogeneous Electrocatalysis},
journal = {Chemical Reviews},
volume = {124},
number = {14},
pages = {8620--8656},
year = {2024},
publisher={ACS Publications}
}

@article{batatia2023mace,
      title={MACE: Higher Order Equivariant Message Passing Neural Networks for Fast and Accurate Force Fields}, 
      author={Ilyes Batatia and Dávid Péter Kovács and Gregor N. C. Simm and Christoph Ortner and Gábor Csányi},
      year={2023},
      journal = {arXiv preprint arXiv:2206.07697v2},
}

@article{sundararaman2022improving,
author = {Sundararaman, Ravishankar and Vigil-Fowler, Derek and Schwarz, Kathleen},
title = {Improving the Accuracy of Atomistic Simulations of the Electrochemical Interface},
journal = {Chemical Reviews},
volume = {122},
number = {12},
pages = {10651-10674},
year = {2022},
}

@article{gross2022abinitio,
author = {Gro{\ss}, Axel and Sakong, Sung},
title = {Ab Initio Simulations of Water/Metal Interfaces},
journal = {Chemical Reviews},
volume = {122},
number = {12},
pages = {10746-10776},
year = {2022},
publisher={ACS Publications}
}

@article{
harraz2025homogeneous,
author = {Deiaa M. Harraz  and Kunal M. Lodaya  and Bryan Y. Tang  and Yogesh Surendranath },
title = {Homogeneous-heterogeneous bifunctionality in Pd-catalyzed vinyl acetate synthesis},
journal = {Science},
volume = {388},
number = {6742},
pages = {eads7913},
year = {2025}}

@article{batzner20223,
  title={E (3)-equivariant graph neural networks for data-efficient and accurate interatomic potentials},
  author={Batzner, Simon and Musaelian, Albert and Sun, Lixin and Geiger, Mario and Mailoa, Jonathan P and Kornbluth, Mordechai and Molinari, Nicola and Smidt, Tess E and Kozinsky, Boris},
  journal={Nature Communications},
  volume={13},
  number={1},
  pages={2453},
  year={2022},
  publisher={Nature Publishing Group UK London}
}

@article{zhuang2025artificial,
  title={An artificial intelligence accelerated ab initio molecular dynamics dataset for electrochemical interfaces},
  author={Zhuang, Yong-Bin and Liu, Chang and Zhu, Jia-Xin and Hu, Jin-Yuan and Le, Jia-Bo and Li, Jie-Qiong and Wen, Xiao-Jian and Fan, Xue-Ting and Jia, Mei and Li, Xiang-Ying and others},
  journal={Scientific Data},
  volume={12},
  number={1},
  pages={997},
  year={2025},
  publisher={Nature Publishing Group UK London}
}

@article{hormann2025machine,
  title={Machine learning and data-driven methods in computational surface and interface science},
  author={H{\"o}rmann, Lukas and Stark, Wojciech G and Maurer, Reinhard J},
  journal={npj Computational Materials},
  volume={11},
  number={1},
  pages={196},
  year={2025},
  publisher={Nature Publishing Group UK London}
}

@article{winther,
  title={Catalysis-Hub. org, an open electronic structure database for surface reactions},
  author={Winther, Kirsten T and Hoffmann, Max J and Boes, Jacob R and Mamun, Osman and Bajdich, Michal and Bligaard, Thomas},
  journal={Scientific Data},
  volume={6},
  number={1},
  pages={75},
  year={2019},
  publisher={Nature Publishing Group UK London}
}

@article{tran2018active,
  title={Active learning across intermetallics to guide discovery of electrocatalysts for CO2 reduction and H2 evolution},
  author={Tran, Kevin and Ulissi, Zachary W},
  journal={Nature Catalysis},
  volume={1},
  number={9},
  pages={696--703},
  year={2018},
  publisher={Nature Publishing Group UK London}
}

@article{Kresse1994,
author = {Kresse, Georg and Hafner, J{\"{u}}rgen},
journal = {Physical Review B},
number = {20},
pages = {14251--14269},
pmid = {10010505},
title = {{Ab initio molecular-dynamics simulation of the liquid-metal–amorphous-semiconductor transition in germanium}},
volume = {49},
year = {1994}
}

@article{Kresse1996a,
author = {Kresse, Georg and Furthm{\"{u}}ller, J{\"{u}}rgen},
journal = {Physical Review B},
number = {16},
pages = {11169--11186},
pmid = {9984901},
title = {{Efficient iterative schemes for ab initio total-energy calculations using a plane-wave basis set}},
volume = {54},
year = {1996}
}

@article{kresse1999ultrasoft,
  title={From ultrasoft pseudopotentials to the projector augmented-wave method},
  author={Kresse, Georg and Joubert, Daniel},
  journal={Physical Review B},
  volume={59},
  number={3},
  pages={1758},
  year={1999},
  publisher={APS}
}

@article{Kresse1996,
author = {Kresse, Georg and Furthm{\"{u}}ller, J{\"{u}}rgen},
journal = {Computational Materials Science},
number = {1},
pages = {15--50},
pmid = {9984901},
title = {{Efficiency of ab-initio total energy calculations for metals and semiconductors using a plane-wave basis set}},
volume = {6},
year = {1996}
}

@article{saleheen2023understanding,
  title={Understanding the influence of solvents on the Pt-catalyzed hydrodeoxygenation of guaiacol},
  author={Saleheen, Mohammad and Mamun, Osman and Verma, Anand Mohan and Sahsah, Dia and Heyden, Andreas},
  journal={Journal of Catalysis},
  volume={425},
  pages={212--232},
  year={2023},
  publisher={Elsevier}
}

@article{klemm2023impact,
  title={Impact of hydrogen bonds on CO2 binding in eutectic solvents: an experimental and computational study toward sorbent design for CO2 capture},
  author={Klemm, Aidan and Vicchio, Stephen P and Bhattacharjee, Sanchari and Cagli, Eda and Park, Yensil and Zeeshan, Muhammad and Dikki, Ruth and Liu, Harrison and Kidder, Michelle K and Getman, Rachel B and others},
  journal={ACS Sustainable Chemistry \& Engineering},
  volume={11},
  number={9},
  pages={3740--3749},
  year={2023},
  publisher={ACS Publications}
}

@article{zhang2020method,
  title={A method for obtaining liquid--solid adsorption rates from molecular dynamics simulations: applied to methanol on Pt (111) in H$_2$O},
  author={Zhang, Xiaohong and Savara, Aditya and Getman, Rachel B},
  journal={Journal of Chemical Theory and Computation},
  volume={16},
  number={4},
  pages={2680--2691},
  year={2020},
  publisher={ACS Publications}
}

@article{kundu2025liquid,
  title={Liquid Phase Modeling in Porous Media: Adsorption of Methanol and Ethanol in H-MFI in Condensed Water},
  author={Kundu, Subrata Kumar and Zeeshan, Muhammad and Watthaisong, Panuwat and Heyden, Andreas},
  journal={Journal of Chemical Theory and Computation},
  year={2025},
  volume = {21},
  number = {12},
  pages = {6121--6134},
  publisher={ACS Publications}
}

@article{grimme2010consistent,
    author = {Grimme, Stefan and Antony, Jens and Ehrlich, Stephan and Krieg, Helge},
    title = {A consistent and accurate ab initio parametrization of density functional dispersion correction (DFT-D) for the 94 elements H-Pu},
    journal = {The Journal of Chemical Physics},
    volume = {132},
    number = {15},
    pages = {154104},
    year = {2010},
    publisher={AIP Publishing}


}

@misc{aqcat25,
  author       = {SandboxAQ},
  title        = {AQCat25 Dataset},
  howpublished = {\url{https://huggingface.co/datasets/SandboxAQ/aqcat25}},
  year         = {2025},
}

@article{opes,
  title={Rethinking metadynamics: from bias potentials to probability distributions},
  author={Invernizzi, Michele and Parrinello, Michele},
  journal={The journal of physical chemistry letters},
  volume={11},
  number={7},
  pages={2731--2736},
  year={2020},
  publisher={ACS Publications}
}

@article{plumed,
  title={Promoting transparency and reproducibility in enhanced molecular simulations},
  journal={Nature methods},
  volume={16},
  number={8},
  pages={670--673},
  year={2019},
  publisher={Nature Publishing Group US New York}
}

@article{ase,
  doi = {10.1088/1361-648x/aa680e},
  url = {https://doi.org/10.1088/1361-648x/aa680e},
  year = 2017,
  month = jun,
  publisher = {{IOP} Publishing},
  volume = {29},
  number = {27},
  pages = {273002},
  author = {Ask Hjorth Larsen and Jens J{\o}rgen Mortensen and Jakob Blomqvist and Ivano E Castelli and Rune Christensen and Marcin Du{\l}ak and Jesper Friis and Michael N Groves and Bj{\o}rk Hammer and Cory Hargus and Eric D Hermes and Paul C Jennings and Peter B Jensen and Adam Kloster and Jens R Kitchin and Mattis Kolsbjerg and Asbj{\o}rn K{\o}rner and Per K Krogh and Fabio Deady Ferrero and Emsel C O F{\aa}rch and S S S and M M and S S S and M M},
  title = {The atomic simulation environment—a Python library for working with atoms},
  journal = {Journal of Physics: Condensed Matter}
}

@article{moon2025catbench,
  title={CatBench framework for benchmarking machine learning interatomic potentials in adsorption energy predictions for heterogeneous catalysis},
  author={Moon, Jinuk and Jeon, Uchan and Choung, Seokhyun and Han, Jeong Woo},
  journal={Cell Reports Physical Science},
  volume={6},
  number={12},
  year={2025},
  publisher={Elsevier}
}

@article{morales2024electrochemical,
  title={Electrochemical CO2 Reduction Mechanism on Copper: Relation between Mesoscopic Mass Transport and Intrinsic Kinetics},
  journal={ResearchSquare},
  author={Morales-Guio, Carlos and Jang, Joonbaek and Ruscher, Martina and Winzely, Maximilian and Rodriguez, Dolores and Reyes-Lopez, Eber and Srivastava, Samanvaya and Christofides, Panagiotis and Sautet, Philippe},
  year={2024}
}

@article{resasco2017promoter,
  title={Promoter effects of alkali metal cations on the electrochemical reduction of carbon dioxide},
  author={Resasco, Joaquin and Chen, Leanne D and Clark, Ezra and Tsai, Charlie and Hahn, Christopher and Jaramillo, Thomas F and Chan, Karen and Bell, Alexis T},
  journal={Journal of the American Chemical Society},
  volume={139},
  number={32},
  pages={11277--11287},
  year={2017},
  publisher={ACS Publications}
}

@article{ringe2019understanding,
  title={Understanding cation effects in electrochemical CO 2 reduction},
  author={Ringe, Stefan and Clark, Ezra L and Resasco, Joaquin and Walton, Amber and Seger, Brian and Bell, Alexis T and Chan, Karen},
  journal={Energy \& Environmental Science},
  volume={12},
  number={10},
  pages={3001--3014},
  year={2019},
  publisher={Royal Society of Chemistry}
}

@article{resasco2025universal,
  title={A Universal Model of Cation Effects in Electrocatalysis},
  author={Resasco, Joaquin},
  journal={JACS Au},
  volume={5},
  number={11},
  pages={5253--5266},
  year={2025},
  publisher={ACS Publications}
}

@article{dominguez2024metal,
  title={Metal--water interface formation: Thermodynamics from ab initio molecular dynamics simulations},
  author={Dom{\'\i}nguez-Flores, Fabiola and Kiljunen, Toni and Gro{\ss}, Axel and Sakong, Sung and Melander, Marko M},
  journal={The Journal of Chemical Physics},
  volume={161},
  number={4},
  year={2024},
  publisher={AIP Publishing}
}

@article{kristoffersen2021towards,
  title={Towards constant potential modeling of CO-CO coupling at liquid water-Cu (1 0 0) interfaces},
  author={Kristoffersen, Henrik H and Chan, Karen},
  journal={Journal of Catalysis},
  volume={396},
  pages={251--260},
  year={2021},
  publisher={Elsevier}
}

@article{kastlunger2022pH,
author = {Kastlunger, Georg and Wang, Lei and Govindarajan, Nitish and Heenen, Hendrik H. and Ringe, Stefan and Jaramillo, Thomas and Hahn, Christopher and Chan, Karen},
title = {Using pH Dependence to Understand Mechanisms in Electrochemical CO Reduction},
journal = {ACS Catalysis},
volume = {12},
number = {8},
pages = {4344-4357},
year = {2022},
doi = {10.1021/acscatal.1c05520},

URL = { 
    
        https://doi.org/10.1021/acscatal.1c05520
    
    

},
eprint = { 
    
        https://doi.org/10.1021/acscatal.1c05520
    
    

}




}

@article{
hahn2017engineering,
author = {Christopher Hahn  and Toru Hatsukade  and Youn-Geun Kim  and Arturas Vailionis  and Jack H. Baricuatro  and Drew C. Higgins  and Stephanie A. Nitopi  and Manuel P. Soriaga  and Thomas F. Jaramillo },
title = {Engineering Cu surfaces for the electrocatalytic conversion of CO<sub>2</sub>: Controlling selectivity toward oxygenates and hydrocarbons},
journal = {Proceedings of the National Academy of Sciences},
volume = {114},
number = {23},
pages = {5918-5923},
year = {2017},
doi = {10.1073/pnas.1618935114},
URL = {https://www.pnas.org/doi/abs/10.1073/pnas.1618935114},
eprint = {https://www.pnas.org/doi/pdf/10.1073/pnas.1618935114}}

@article{amirbeigiarab2023atomic,
  author  = {Amirbeigiarab, Reihaneh and Tian, Jing and Herzog, Antonia and Qiu, Canrong and Bergmann, Arno and Roldan Cuenya, Beatriz and Magnussen, Olaf M.},
  title   = {Atomic-scale surface restructuring of copper electrodes under CO2 electroreduction conditions},
  journal = {Nature Catalysis},
  volume  = {6},
  pages   = {837--846},
  year    = {2023},
  doi     = {10.1038/s41929-023-01009-z},
  url     = {https://doi.org/10.1038/s41929-023-01009-z}
}

@article{gauthier2019unified,
  title={Unified approach to implicit and explicit solvent simulations of electrochemical reaction energetics},
  author={Gauthier, Joseph A and Dickens, Colin F and Heenen, Hendrik H and Vijay, Sudarshan and Ringe, Stefan and Chan, Karen},
  journal={Journal of chemical theory and computation},
  volume={15},
  number={12},
  pages={6895--6906},
  year={2019},
  publisher={ACS Publications}
}

@article{nguyen2024influence,
  title={The influence of mesoscopic surface structure on the electrocatalytic selectivity of CO2 reduction with UHV-prepared Cu (111) single crystals},
  author={Nguyen, Khanh-Ly C and Bruce, Jared P and Yoon, Aram and Navarro, Juan J and Scholten, Fabian and Landwehr, Felix and Rettenmaier, Clara and Heyde, Markus and Cuenya, Beatriz Roldan},
  journal={ACS Energy Letters},
  volume={9},
  number={2},
  pages={644--652},
  year={2024},
  publisher={ACS Publications}
}

@article{hori2008electrochemical,
  title={Electrochemical CO2 reduction on metal electrodes},
  author={Hori, Yoshio},
  journal={Modern aspects of electrochemistry},
  pages={89--189},
  year={2008},
  publisher={Springer}
}

@article{nitopi2019progress,
  title={Progress and perspectives of electrochemical CO2 reduction on copper in aqueous electrolyte},
  author={Nitopi, Stephanie and Bertheussen, Erlend and Scott, Soren B and Liu, Xinyan and Engstfeld, Albert K and Horch, Sebastian and Seger, Brian and Stephens, Ifan EL and Chan, Karen and Hahn, Christopher and others},
  journal={Chemical reviews},
  volume={119},
  number={12},
  pages={7610--7672},
  year={2019},
  publisher={ACS Publications}
}

@article{garza2018mechanism,
  title={Mechanism of CO2 reduction at copper surfaces: pathways to C2 products},
  author={Garza, Alejandro J and Bell, Alexis T and Head-Gordon, Martin},
  journal={Acs Catalysis},
  volume={8},
  number={2},
  pages={1490--1499},
  year={2018},
  publisher={ACS Publications}
}

@article{goodpaster2016identification,
  title={Identification of possible pathways for C--C bond formation during electrochemical reduction of CO2: new theoretical insights from an improved electrochemical model},
  author={Goodpaster, Jason D and Bell, Alexis T and Head-Gordon, Martin},
  journal={The journal of physical chemistry letters},
  volume={7},
  number={8},
  pages={1471--1477},
  year={2016},
  publisher={ACS Publications}
}

@article{perez2017structure,
  title={Structure-and potential-dependent cation effects on CO reduction at copper single-crystal electrodes},
  author={P{\'e}rez-Gallent, Elena and Marcandalli, Giulia and Figueiredo, Marta Costa and Calle-Vallejo, Federico and Koper, Marc TM},
  journal={Journal of the American Chemical Society},
  volume={139},
  number={45},
  pages={16412--16419},
  year={2017},
  publisher={ACS Publications}
}

@article{vineyard1957frequency,
  title={Frequency factors and isotope effects in solid state rate processes},
  author={Vineyard, George H},
  journal={Journal of Physics and Chemistry of Solids},
  volume={3},
  number={1-2},
  pages={121--127},
  year={1957},
  publisher={Elsevier}
}

@article{invernizzi2021opes,
  title={OPES: On-the-fly probability enhanced sampling method},
  author={Invernizzi, Michele},
  journal={arXiv preprint arXiv:2101.06991},
  year={2021}
}

@article{islam2023implicit,
  title={An implicit electrolyte model for plane wave density functional theory exhibiting nonlinear response and a nonlocal cavity definition},
  author={Islam, SM and Khezeli, Foroogh and Ringe, Stefan and Plaisance, Craig},
  journal={The Journal of Chemical Physics},
  volume={159},
  number={23},
  year={2023},
  publisher={AIP Publishing}
}

@article{henkelman1999dimer,
  title={A dimer method for finding saddle points on high dimensional potential surfaces using only first derivatives},
  author={Henkelman, Graeme and J{\'o}nsson, Hannes},
  journal={The Journal of chemical physics},
  volume={111},
  number={15},
  pages={7010--7022},
  year={1999},
  publisher={American Institute of Physics}
}

@article{liu2021promotional,
  title={Promotional role of a cation intermediate complex in C2 formation from electrochemical reduction of CO2 over Cu},
  author={Liu, Hong and Liu, Jian and Yang, Bo},
  journal={ACS Catalysis},
  volume={11},
  number={19},
  pages={12336--12343},
  year={2021},
  publisher={ACS Publications}
}

@article{scholten2021identifying,
  title={Identifying Structure--Selectivity Correlations in the Electrochemical Reduction of CO2: A Comparison of Well-Ordered Atomically Clean and Chemically Etched Copper Single-Crystal Surfaces},
  author={Scholten, Fabian and Nguyen, Khanh-Ly C and Bruce, Jared P and Heyde, Markus and Roldan Cuenya, Beatriz},
  journal={Angewandte Chemie International Edition},
  volume={60},
  number={35},
  pages={19169--19175},
  year={2021},
  publisher={Wiley Online Library}
}

@article{gauthier2021role,
  title={The role of roughening to enhance selectivity to C2+ products during CO2 electroreduction on copper},
  author={Gauthier, Joseph A and Stenlid, Joakim Halldin and Abild-Pedersen, Frank and Head-Gordon, Martin and Bell, Alexis T},
  journal={ACS Energy Letters},
  volume={6},
  number={9},
  pages={3252--3260},
  year={2021},
  publisher={ACS Publications}
}

@article{zhang2023promoting,
  title={Promoting Cu-catalysed CO2 electroreduction to multicarbon products by tuning the activity of H2O},
  author={Zhang, Hao and Gao, Jiaxin and Raciti, David and Hall, Anthony Shoji},
  journal={Nature Catalysis},
  volume={6},
  number={9},
  pages={807--817},
  year={2023},
  publisher={Nature Publishing Group UK London}
}

@article{montoya2015theoretical,
  title={Theoretical insights into a CO dimerization mechanism in CO2 electroreduction},
  author={Montoya, Joseph H and Shi, Chuan and Chan, Karen and N{\o}rskov, Jens K},
  journal={The journal of physical chemistry letters},
  volume={6},
  number={11},
  pages={2032--2037},
  year={2015},
  publisher={ACS Publications}
}

@article{cheng2017full,
  title={Full atomistic reaction mechanism with kinetics for CO reduction on Cu (100) from ab initio molecular dynamics free-energy calculations at 298 K},
  author={Cheng, Tao and Xiao, Hai and Goddard III, William A},
  journal={Proceedings of the National Academy of Sciences},
  volume={114},
  number={8},
  pages={1795--1800},
  year={2017},
  publisher={National Academy of Sciences}
}

\clearpage
\newpage

\let\addcontentsline\oldaddcontentsline
\setcounter{figure}{0}
\beginappendix
\renewcommand{\figurename}{Supplementary Figure}
\renewcommand{\tablename}{Supplementary Table}
\section{OC25 dataset}

\subsection{OC25 distribution compared to OC20 and OC22}
\begin{figure}[h!]
    \centering
    \includegraphics[width=\linewidth]{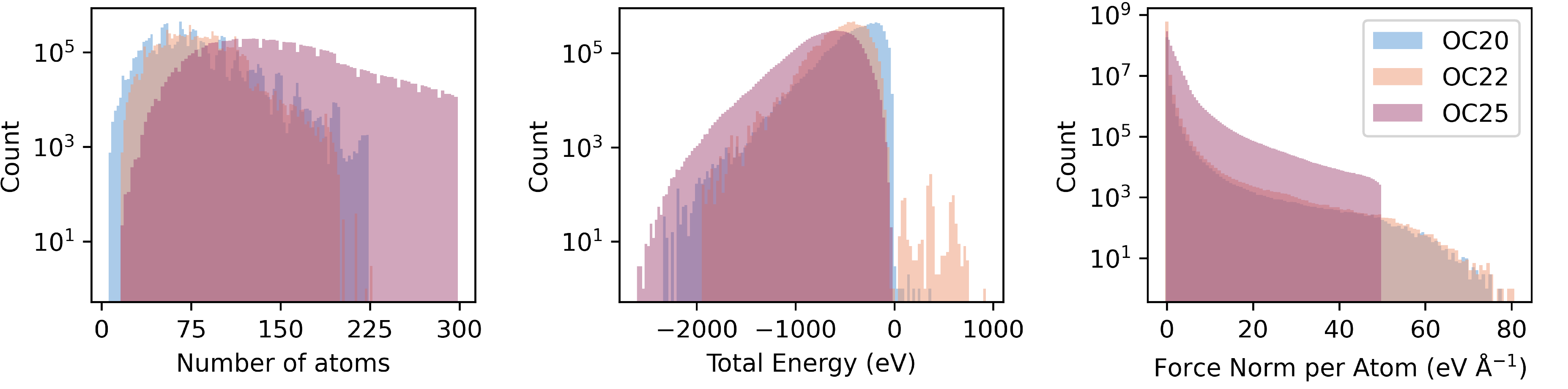}
    \caption{Distribution of number of atoms, total energy, and force norm across the OC25, OC20, and OC22 datasets.}
    \label{fig:oc25_comp_dist}
\end{figure}

\subsection{Statistics}
\begin{figure}[h!]
    \centering
    \includegraphics[width=\linewidth]{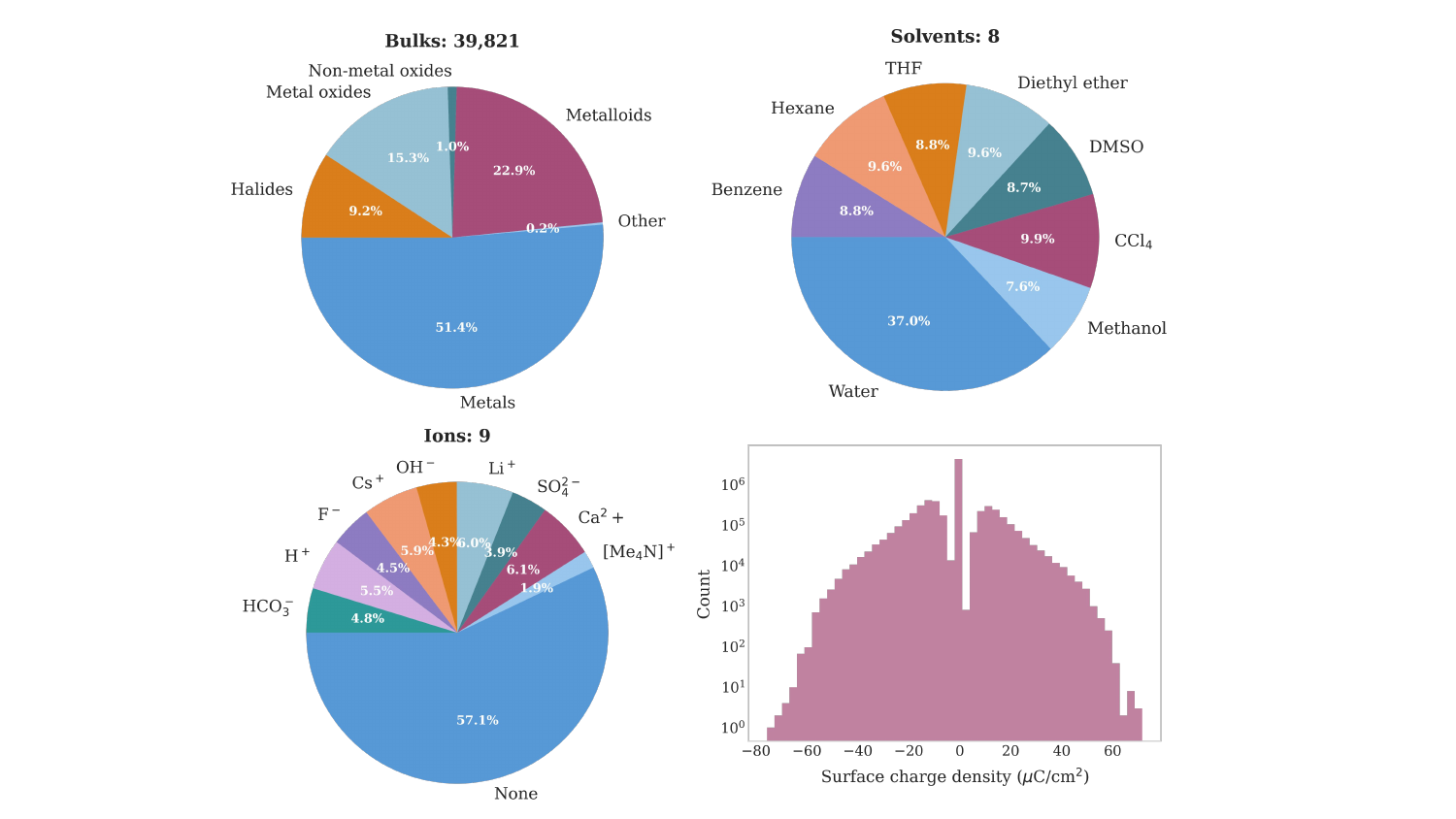}
    \caption{Overview of the bulks, solvents, ions sampled in OC25 and the surface charge distribution (in µC/cm$^2$) for the metallic interfaces in the dataset. Adsorbates are sampled from the same set of OC20, with the addition of a few reactive intermediates.}
    \label{fig:oc25_stats}
\end{figure}

\subsection{Training splits}\label{sec:splits}
Dataset splits are summarized in Supplementary Table \ref{tab:dataset}.
\begin{table}[ht!]
\caption{Size of the \oc ~train, validation, and test splits.}
\centering
\resizebox{0.45\textwidth}{!}{%
\begin{tabular}{ccrl}
\toprule
\multicolumn{1}{l}{}  & Split   & Size      & \multicolumn{1}{c}{Description}        \\ \midrule
Train & All     & 7,395,512 & Training set \\ \midrule
Val             &
Val   &  203,630  & OOD combos \\ \midrule
\multirow{5}{*}{Test}  &
Test    &  202,119  & OOD combos \\
& Solvent   & 11,111     &   OOD solvents \\
& Ion  & 7,176     &   OOD ions   \\
& Both & 6,989 & OOD solvents+ions \\ 
& Solvation & 5,713 & $\tilde{\Delta E_{solv}}$ \\ \bottomrule
\end{tabular}%
}
\label{tab:dataset}
\end{table}

\subsection{Additional adsorbates}\label{sec:ads}
A total of 98 adsorbates were sampled from to create OC25 configurations. These included all of the OC20 adsorbates\cite{oc20} as well as adsorbates presented in OC20NEB\cite{wander2025cattsunami} and OCx24\cite{abed2024open}. These additional adsorbates are presented in Supplementary Table \ref{tab:adsorbates}. We refer readers to the OC20 paper for their full list of adsorbates.

\begin{table*}[h!]
    \centering
    \caption{Additional adsorbates considered in OC25 alongside the full set of OC20 adsorbates.}
    \begin{tabular}{|m{3cm}|m{11cm}|}
        \hline
        Adsorbate class     &     Adsorbates \\
        \hline & \\
        \ce{O}$/$\ce{H} Only      &
                                \ce{^{*}OOH},
                                \ce{^{*}H2} \\ 
        \hline & \\
        C\textsubscript{1}    &
                                \ce{^{*}OCHO}, 
                                \ce{^{*}COOH},
                                \ce{^{*}OC^{*}O} \\
        \hline & \\
        C\textsubscript{2}    &
                                \ce{CO^{*}COH},
                                \ce{^{*}CCOH},
                                \ce{^{*}CH2CH2^{*}O},
                                \ce{^{*}CHCH2^{*}O},
                                \ce{^{*}COHCH2^{*}O},
                                \ce{^{*}CH2OH^{*}CH2OH},
                                \ce{^{*}OCCHOH},
                                \ce{^{*}OCH2CH2OH},
                                \ce{^{*}OCH2CH2^{*}O},
                                \ce{^{*}OCH2CHO},
                                \ce{O^{*}C^{*}CO} \\
        \hline
    \end{tabular}
    \label{tab:adsorbates}
\end{table*}

\section{Model training}
Baseline model and training hyperparameters followed the same procedures originally proposed in OMol25\cite{omol} and UMA\cite{wood2025family}. Models were trained with the AdamW optimizer\cite{adamw}, a learning rate of 8e-4, and trained for 40 epochs. Direct models followed a multi-stage scheme, first trained on BF16 and then finetuned at FP32 with a learning rate of 4e-4. A per-atom MAE loss was used for energies and a L2-norm loss for forces, with energy and force coefficients of 10. UMA-S-1.1 was taken directly from the publicly released checkpoints at \href{https://huggingface.co/facebook/UMA}{https://huggingface.co/facebook/UMA}. UMA-S-1.1 was finetuned through the ``oc20'' task head, with all 32 experts available. All models were trained on Nvidia H100 80GB GPU cards. Model hyperparameters are provided in Table \ref{tab:hyperparams}.

\begin{table}[h!]
\caption{Training and model hyperparameters for the baseline models trained in this work.}
\centering
\resizebox{0.7\textwidth}{!}{%
\begin{tabular}{l|c|c|c|c}
\toprule
Hyperparameters & eSEN-S-d./cons. & eSEN-M  & UMA-S-ft \\ \midrule
\# sphere channels & 128 & 128 & 128 \\
lmax & 2 & 4  & 2 \\
mmax & 2 & 2 & 2 \\
\# moe experts & 0 & 0 & 32 \\
max neighbors & 30/300 & 30 & 300 \\
cutoff radius & 6 & 6 & 6 \\
\# edge channels & 128 & 128 & 128 \\
distance function & gaussian & gaussian & gaussian \\
\# distance basis & 64 & 128 & 64 \\
\# layers & 4 & 10 & 4 \\
\# hidden channels & 128 & 128 & 128 \\
learning rate & 8e-4 & 8e-4 & 4e-4 \\
\# gpus & 32/64 & 32 & 64 \\
batch size ( \# atoms) & 76800 & 44800 & 76800 \\
energy coeff. & 10 & 10 & 10 \\
force coeff. & 10 & 10 & 10 \\
\# of params. & 6.3M & 50.7M & 146.6M \\
\bottomrule
\end{tabular}
}
\label{tab:hyperparams}
\end{table}
\clearpage

\section{Additional results}
\subsection{Validation}
Baseline model results on the validation set are provided in Table \ref{tab:val-results}.
\begin{table}[h!]
\caption{
Baseline results for the validation split. Energy and force mean absolute errors (MAE) are reported in units of eV and eV/\angs.
}
\label{tab:val-results}
\centering
\resizebox{0.7\textwidth}{!}{%
\begin{tabular}{@{}llrcc@{}}
\toprule
 & &  & \multicolumn{2}{c}{Validation} \\ \midrule
Dataset & Model & \# of params & Energy & Forces \\ \midrule
\multirow{3}{*}{OC25} 
& eSEN-S-d. & 6.3M & 0.138 & 0.020 \\
& eSEN-S-cons. & 6.3M & 0.104 & 0.015 \\
& eSEN-M-d. & 50.7M & 0.061 & 0.009 \\ \midrule
UMA & UMA-S-1.1 & 146.6M & - & 0.064 \\
UMA$\rightarrow$OC25 & UMA-S-ft & 146.6M & 0.093 & 0.014 \\ \bottomrule
\end{tabular}%
}
\end{table}

\subsection{Error by solvent and ion type}
We breakdown model performance as a function of the different solvents and ions. Results are provided for eSEN-S-cons. in Supplementary Figure \ref{fig:breakdown}.

\begin{figure}[h!]
    \centering
    \includegraphics[width=\linewidth]{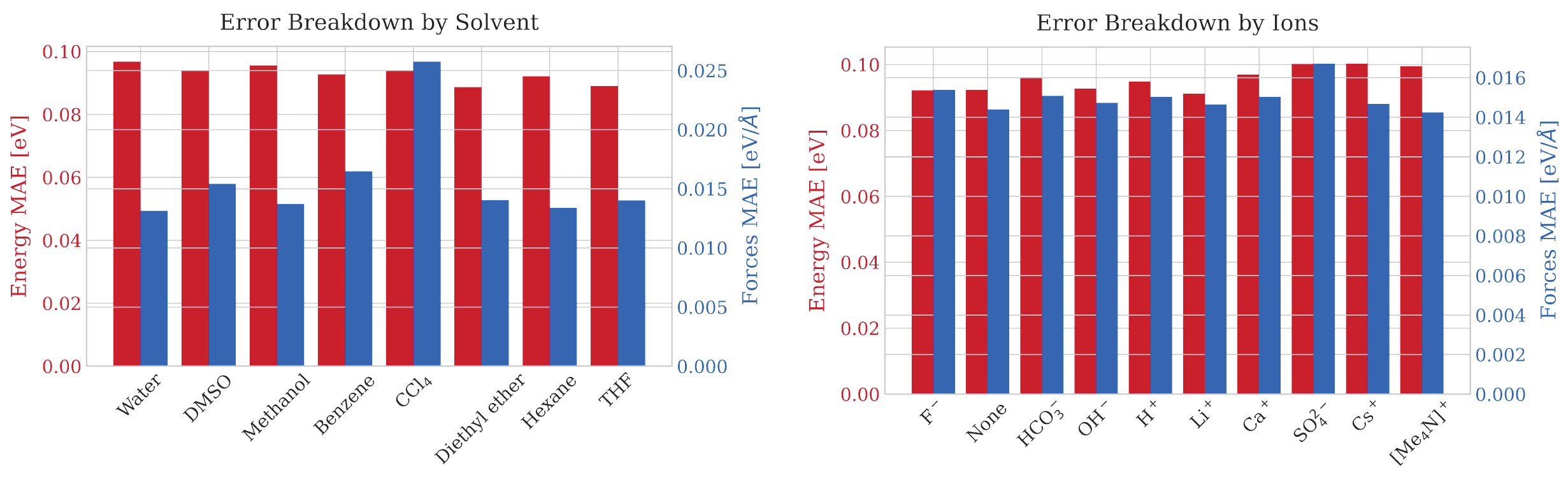}
    \caption{Energy and force mean absolute errors (MAE) broken down for the different solvent and ion types in the validation split. Results reported for the eSEN-S-cons. model.}
    \label{fig:breakdown}
\end{figure}

\subsection{Solvation energy breakdown}\label{sec:breakdown}
For this work, we propose a pseudo solvation energy, $\tilde{\Delta E_{solv}}$, to evaluate baseline models across. Here, we generate a static solid-liquid interface configuration and construct the vacuum and reference configurations directly from this structure (i.e. deleting the solvent to generate the vacuum adsorbate+surface configuration). A break down of the different errors across the terms are provided in Table \ref{tab:solvation}

\begin{equation}
    \tilde{\Delta E_{solv}} = \tilde{\Delta E_{ads}^{solv}} - \tilde{\Delta E_{ads}^{vac}}
\end{equation}

\subsection{Force convergence plots on the filtered set}\label{sec:nodrift}
\begin{figure}
    \centering
    \includegraphics[width=\linewidth]{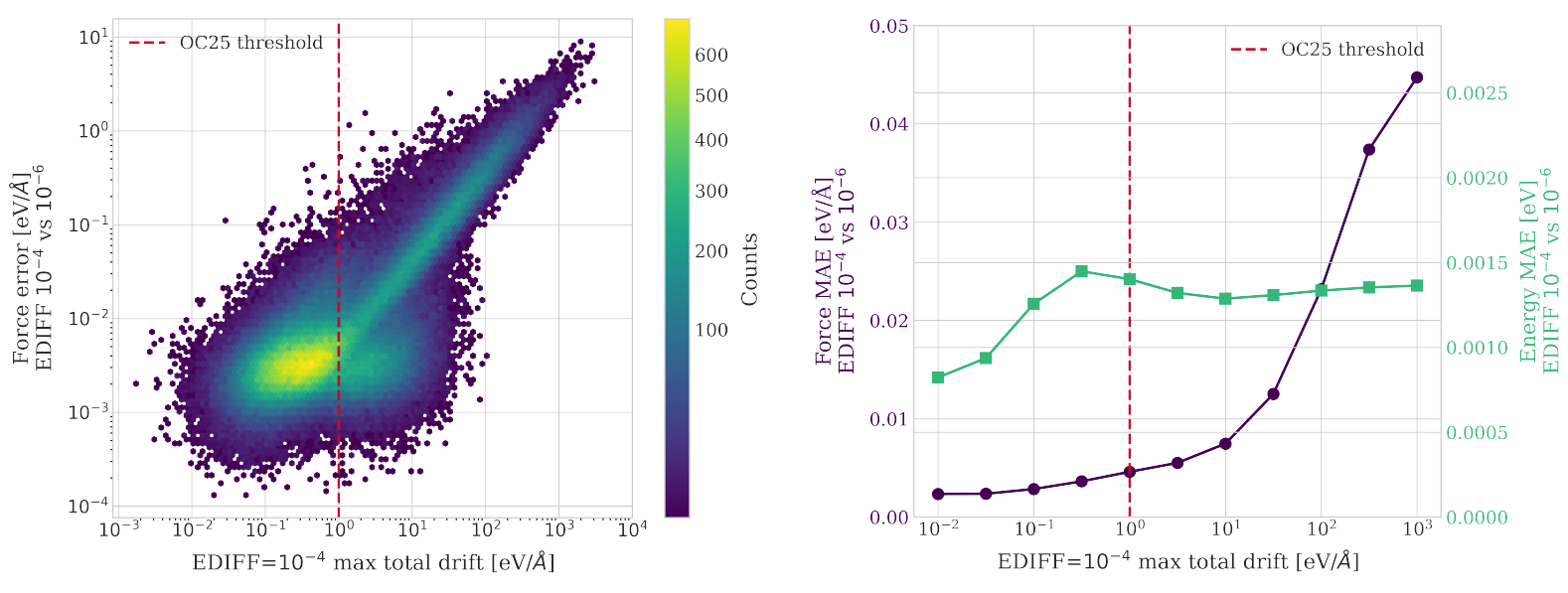}
    \caption{DFT force convergence errors as a function of the total drift in calculations with an electronic termination of $10^{-4}$~eV (``EDIFF''). Errors are computed against more tightly converged ($10^{-6}$~eV) calculations for a $\sim$300k subset of the dataset. A threshold of 1 eV/\angs~on the max drift is selected for the OC25 training dataset. All validation and test sets were calculated with the tighter convergence settings.
    }
    \label{fig:drift}
\end{figure}
\begin{figure}[h!]
    \centering
    \includegraphics[width=\linewidth]{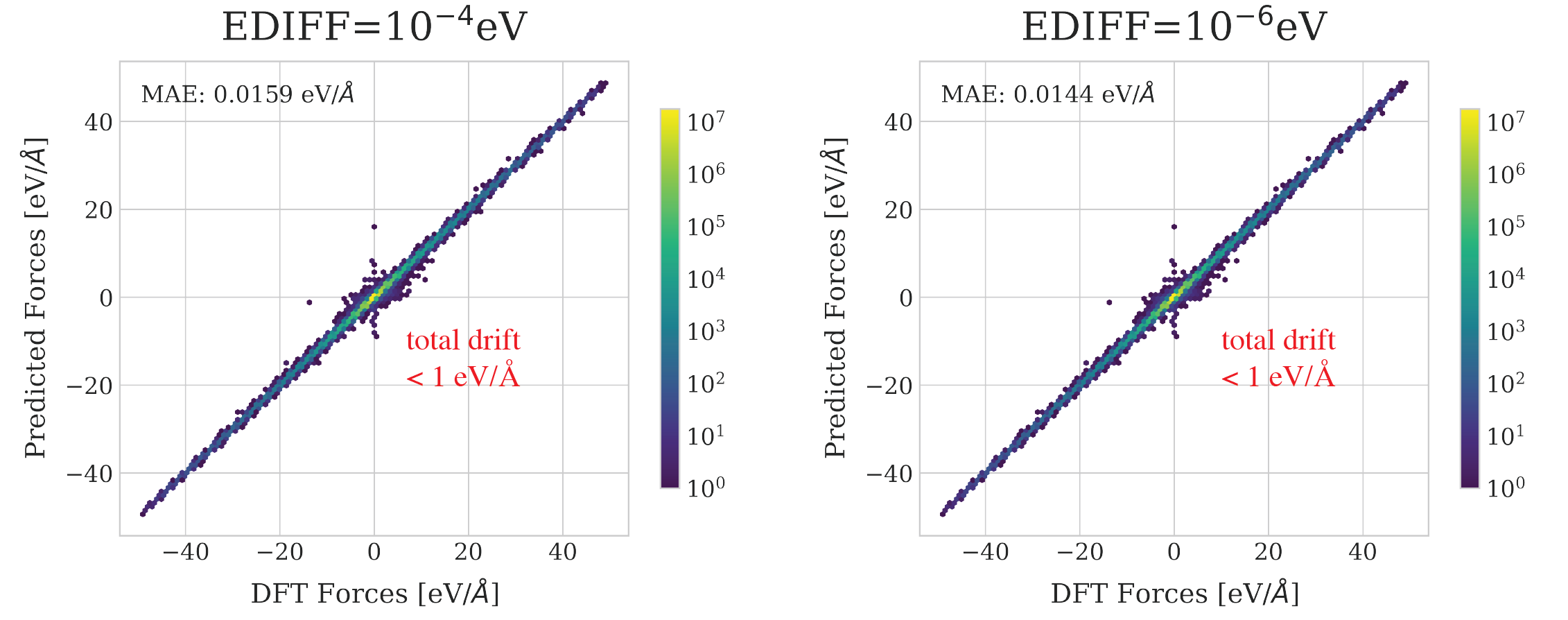}
    \caption{Parity plots of energy and force predictions of OC25 under different evaluation paradigms. A single model is trained on the filtered, released OC25 dataset and evaluated on an identical validation set calculated with the original (EDIFF=$10^{-4}$eV) and tighter (EDIFF=$10^{-6}$eV) settings.}
    \label{fig:parity_nodrift}
\end{figure}

The final OC25 dataset ultimately filtered samples with a total drift <1eV/\angs~to ensure the dataset is broadly useful beyond just MLIP training. Supplementary Figure \ref{fig:parity_nodrift} corresponds to a model trained on the released OC25 dataset (converged with EDIFF=10$^{-4}$ eV, but with all images with total drift >1 eV/\angs~removed) and evaluated on the validation set with the same drift filtering. The small gap between force errors at EDIFF=$10^{-4}$eV) and (EDIFF=$10^{-6}$eV) suggests that the total drift filter does a reasonable job at removing problematic samples in the training set. 

\begin{table}[]
\caption{Baseline solvation energy results broken down across the different components. Where, $\tilde{\Delta E_{solv}}$ is the pseudo solvation energy, $\tilde{\Delta E_{ads}^{solv}}$ is the adsorption energy on the solid-liquid interface, and $\tilde{\Delta E_{ads}^{vac}}$ is the adsorption energy in vacuum.
}
\label{tab:solvation}
\centering
\resizebox{0.7\textwidth}{!}{%
\begin{tabular}{@{}llrccl@{}}
\toprule
 &  &  & \multicolumn{3}{c}{Energy MAE {[}eV{]}} \\ \midrule
Dataset & Model & \# of params & $\tilde{\Delta E_{solv}}$ & $\tilde{\Delta E_{ads}^{solv}}$ & $\tilde{\Delta E_{ads}^{vac}}$ \\ \midrule
\multirow{3}{*}{OC25} & eSEN-S-d. & 6.3M & 0.060 & 0.071 & 0.075 \\
 & eSEN-S-cons. & 6.3M & 0.045 & 0.057 & 0.057 \\
 & eSEN-M-d. & 50.7M & 0.040 & 0.041 & 0.050 \\ \midrule
UMA & UMA-S-1.1 & 146.6M & 0.169 & 0.520 & 0.407 \\
UMA$\rightarrow$OC25 & UMA-S-ft & 146.6M & 0.136 & 0.053 & 0.147 \\ \bottomrule
\end{tabular}%
}
\end{table}

\section{Benchmarking OC25-trained models against AIMD reference simulations}
\subsection{Oxygen and hydrogen density profiles for metal/water interfaces}

\begin{figure}[H]
    \centering
    \includegraphics[width=0.6\linewidth]{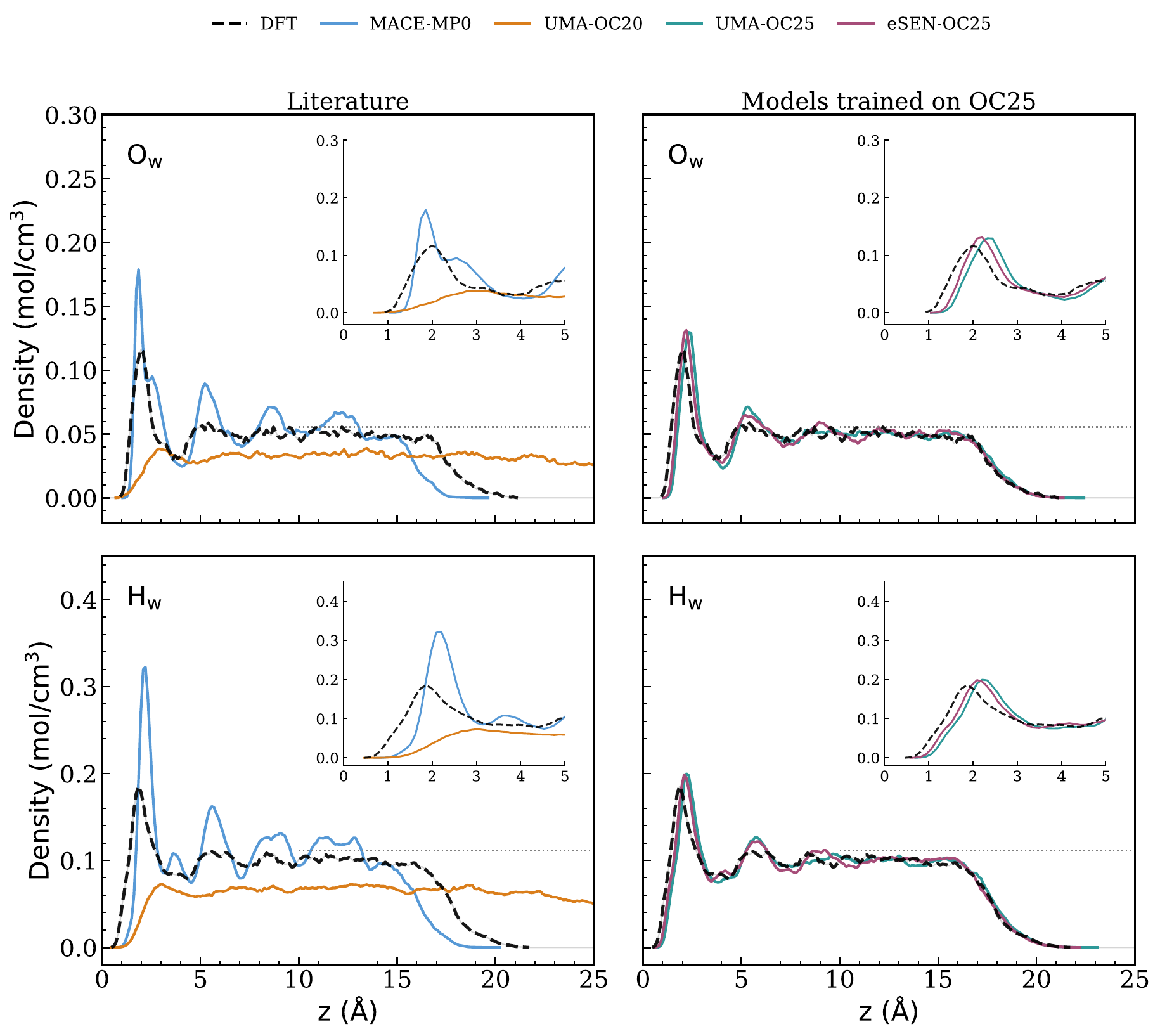}
    \caption{Density profiles for Ag111.}
    \label{fig:si:Ag111}
\end{figure}

\begin{figure}[H]
    \centering
    \includegraphics[width=0.6\linewidth]{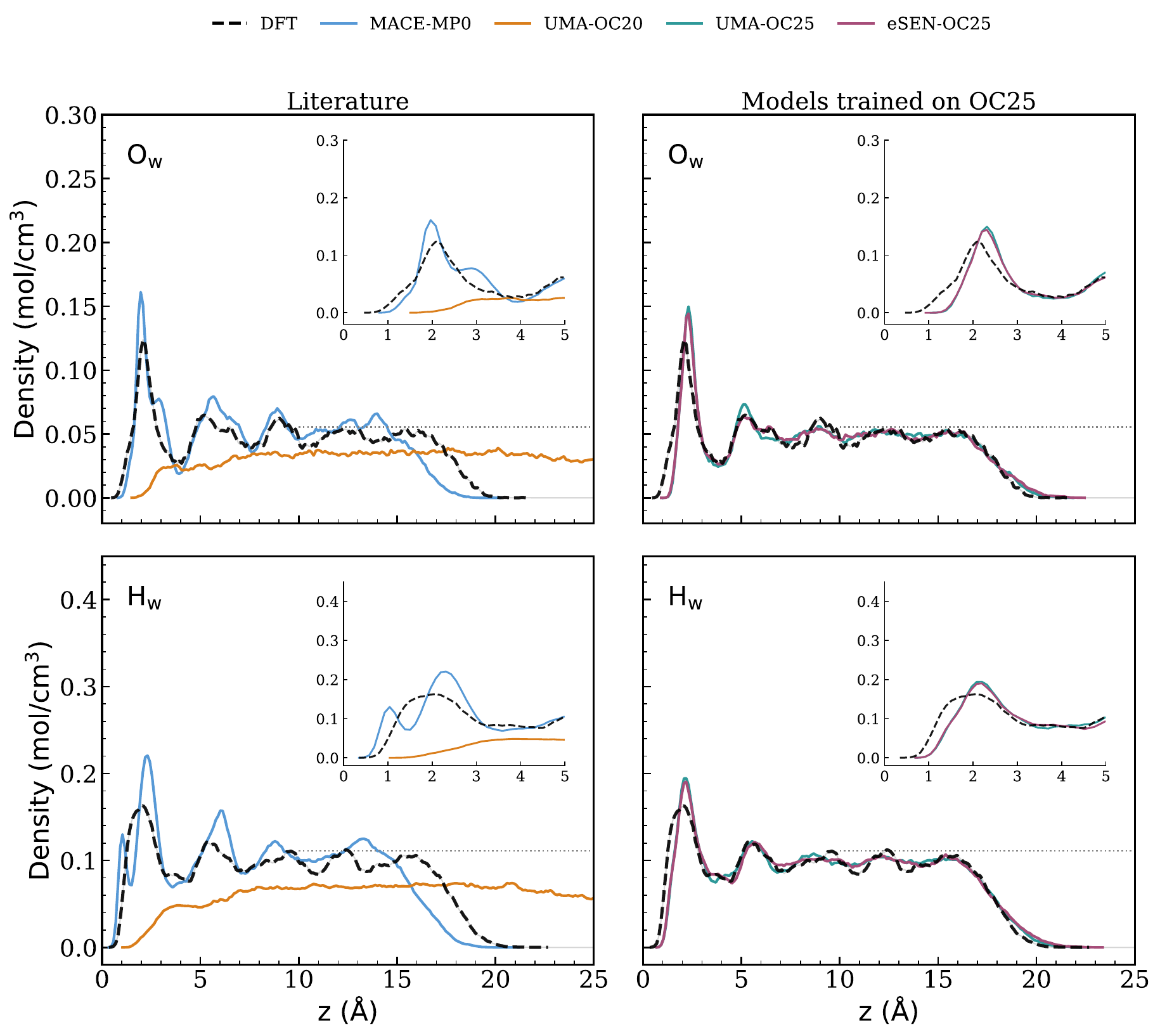}
    \caption{Density profiles for Au111.}
    \label{fig:si:Au111}
\end{figure}

\begin{figure}[H]
    \centering
    \includegraphics[width=0.6\linewidth]{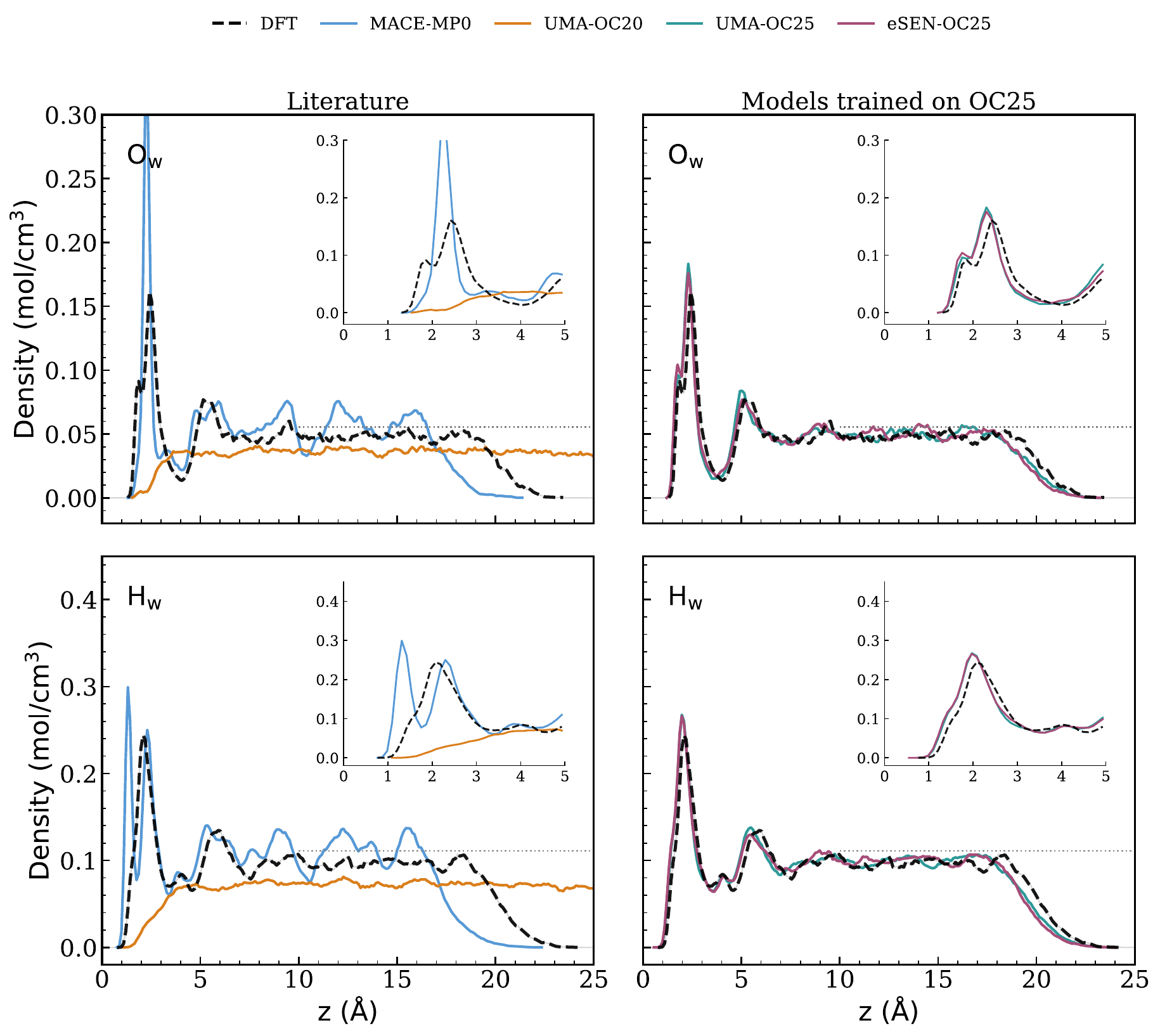}
    \caption{Density profiles for Pd111.}
    \label{fig:si:Pd111}
\end{figure}

\begin{figure}[H]
    \centering
    \includegraphics[width=0.6\linewidth]{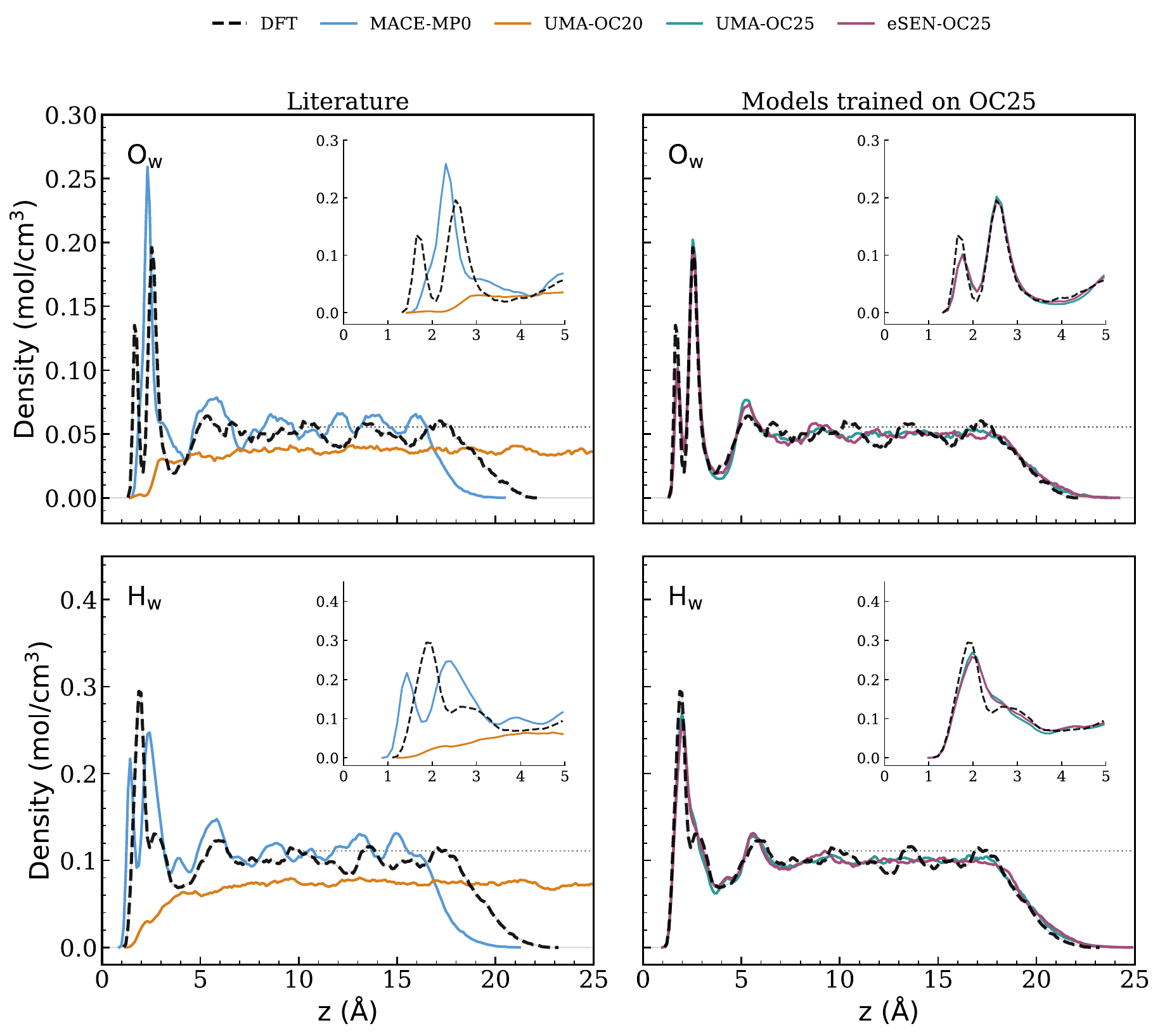}
    \caption{Density profiles for Pt111.}
    \label{fig:si:Pt111}
\end{figure}

\begin{figure}[H]
    \centering
    \includegraphics[width=0.6\linewidth]{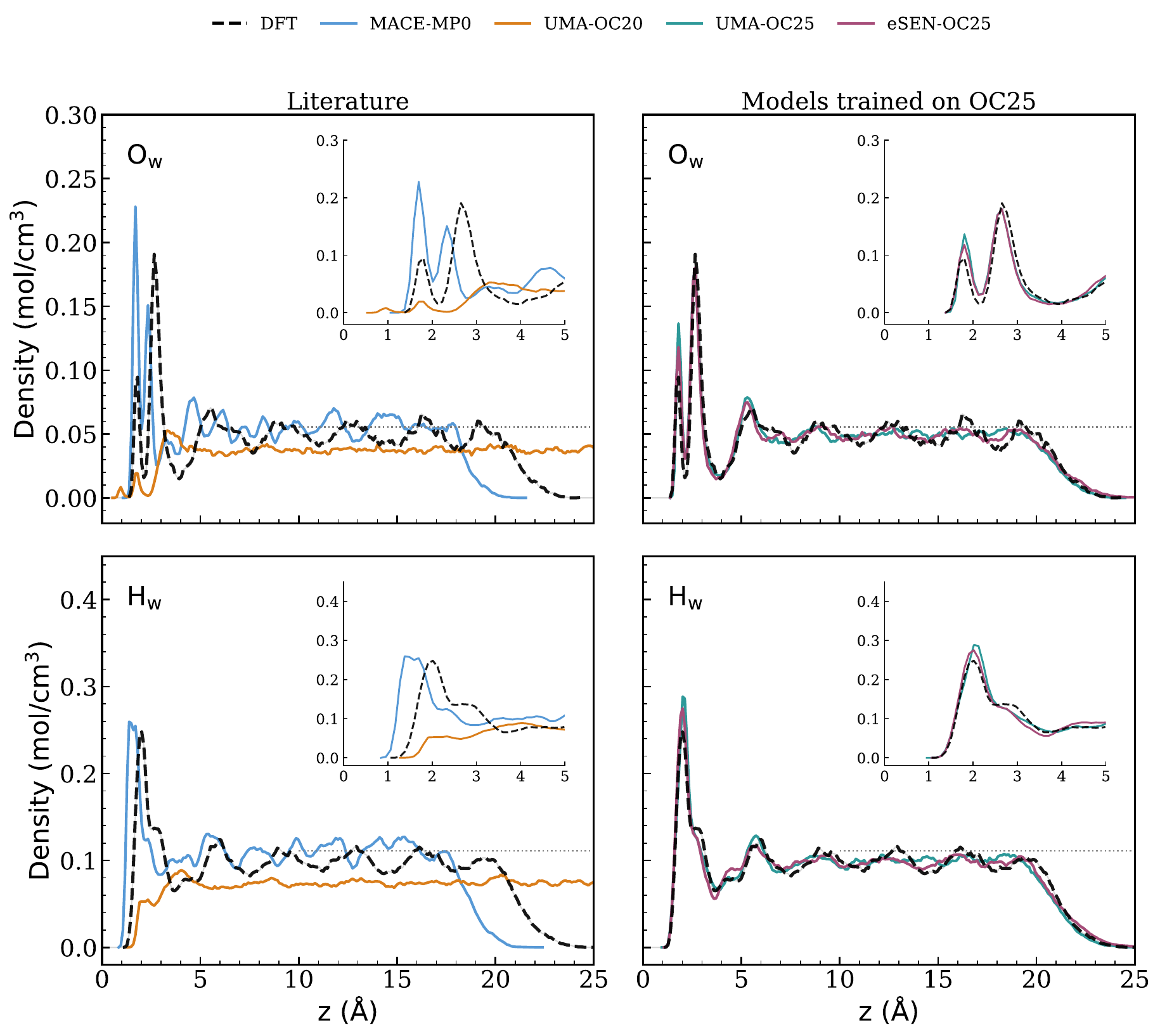}
    \caption{Density profiles for Rh111.}
    \label{fig:si:Rh111}
\end{figure}

\begin{figure}[H]
    \centering
    \includegraphics[width=0.6\linewidth]{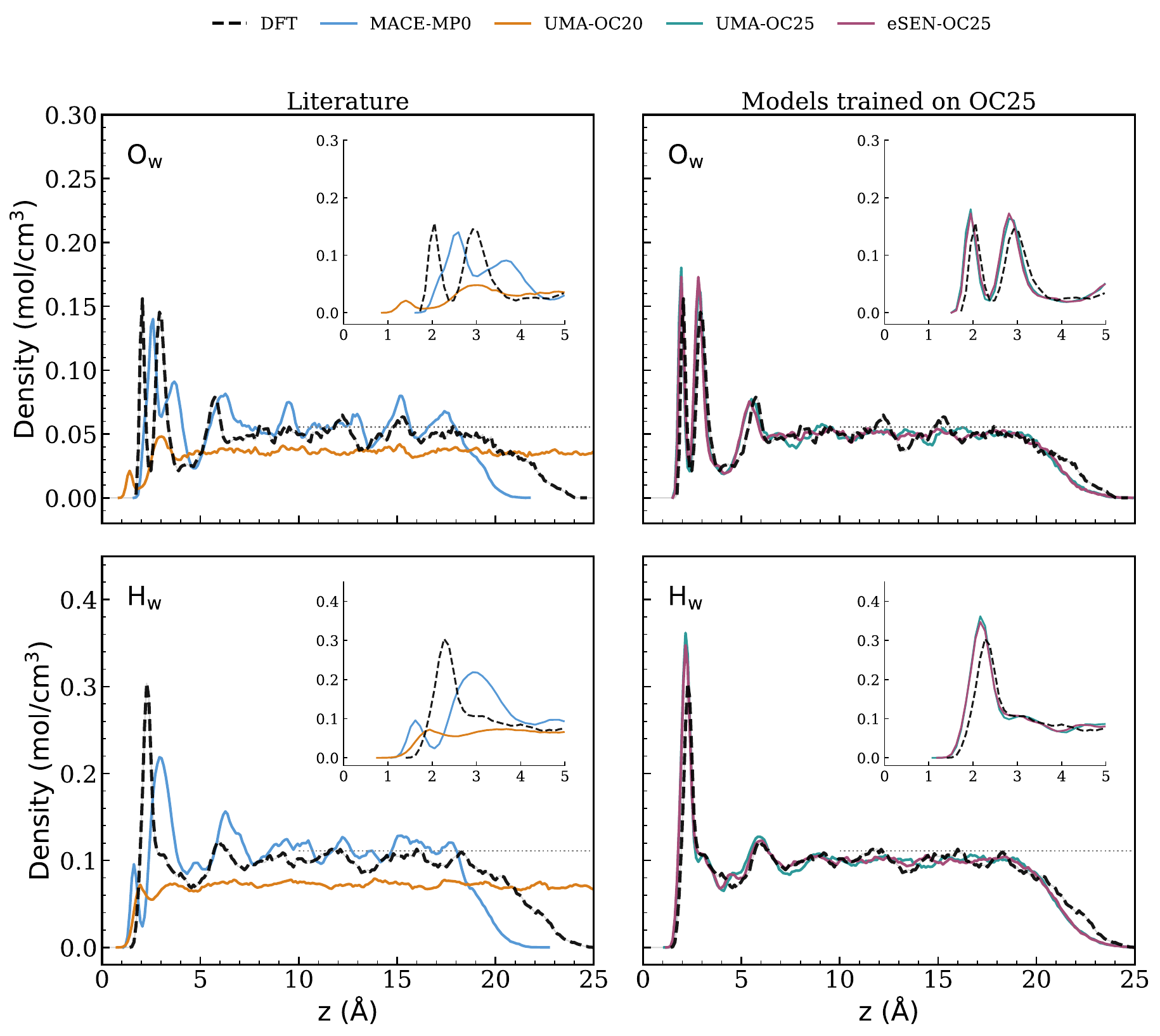}
    \caption{Density profiles for Ru0001.}
    \label{fig:si:Ru0001}
\end{figure}

\subsection{CO dimerization on a 3$\times$4 Cu(100)/water interface at neural and 2 (Cs$^{+}$ + e$^{-}$)}
\begin{figure}[H]
    \centering
    \includegraphics[width=0.79\linewidth]{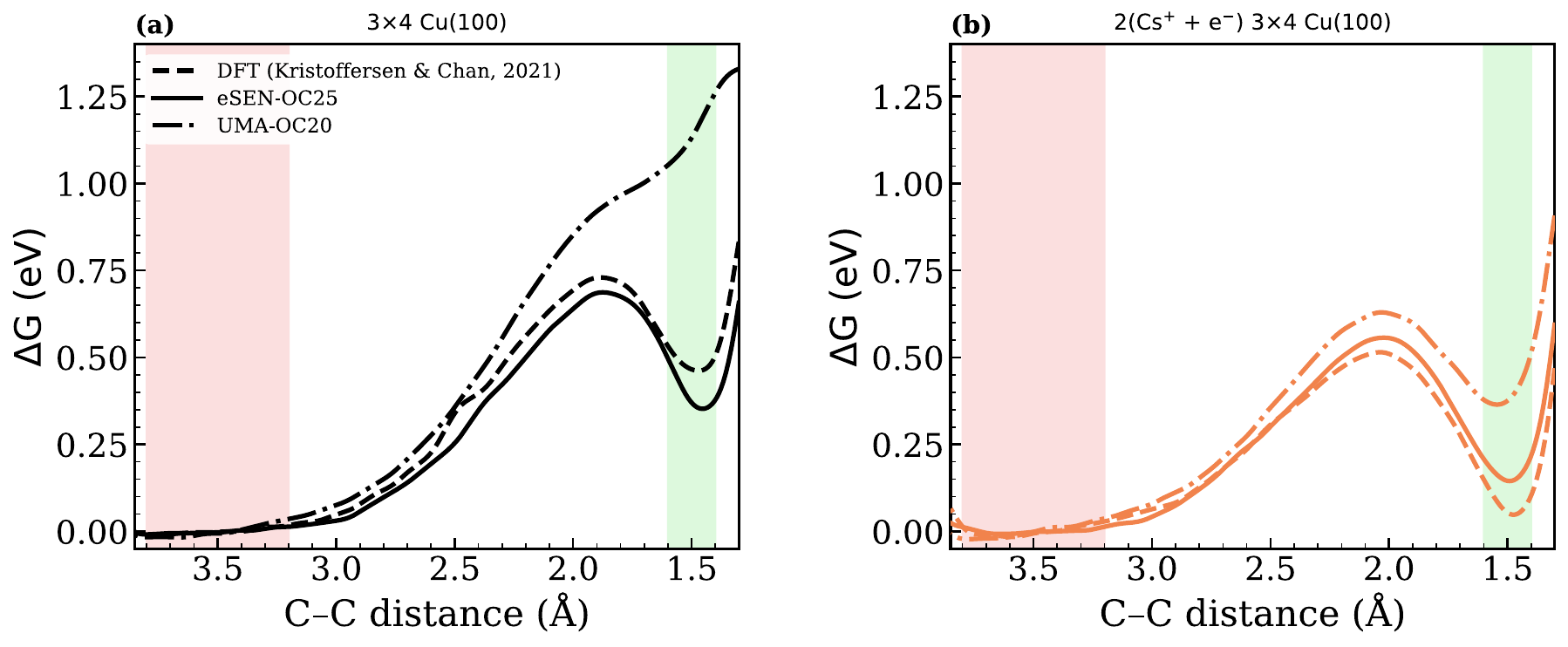}
    \caption{Free energy profile for CO dimerization as a function of C-C distance comparing eSEN-OC25 (solid), UMA-OC20 (dashdot), and AIMD reference data (dashed).}
    \label{fig:si:Ru0001}
\end{figure}

\newpage
\section{CO dimerization on Cu(100)}

\subsection{Effect of surface charging and cation identity}
\begin{figure}[H]
    \centering
    \includegraphics[width=0.85\linewidth]{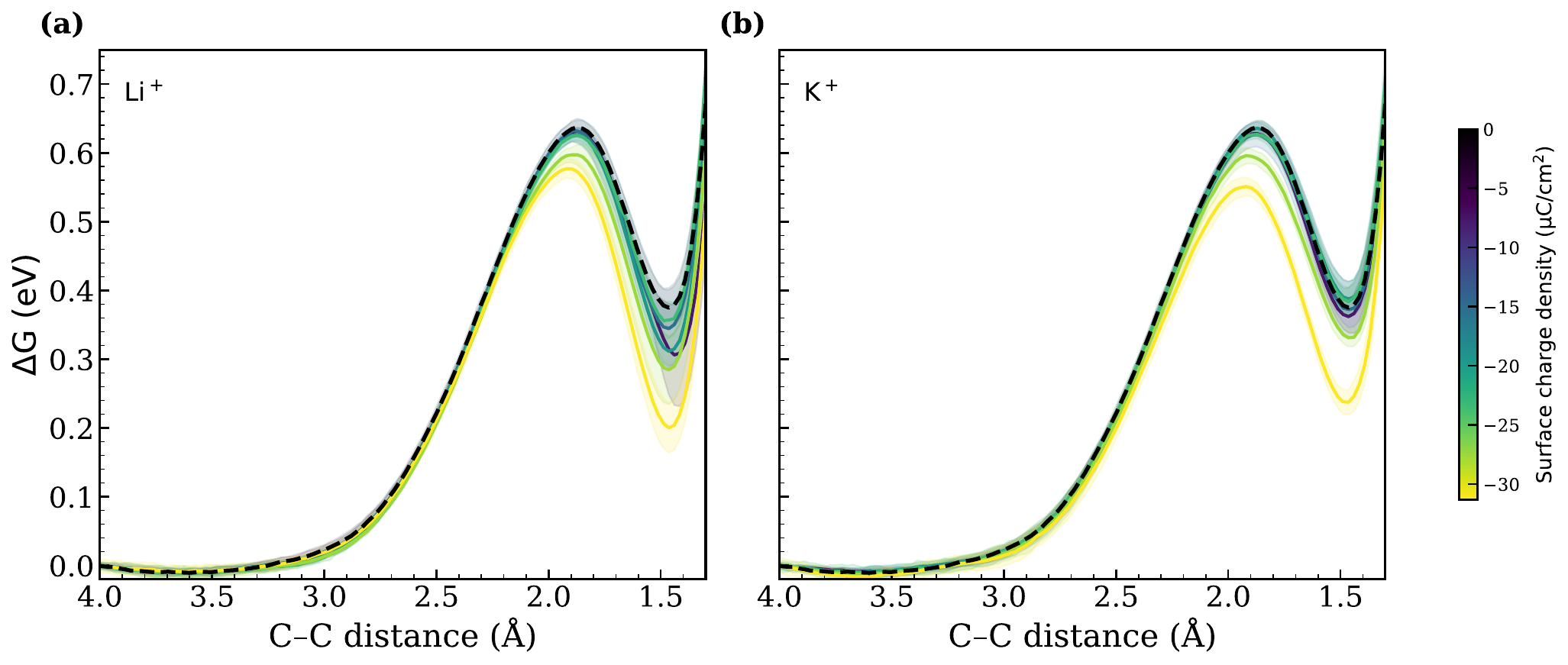}
    \caption{Free energy profiles along the C-C distance collective variable as a function of interfacial \textbf{(a)} Li$^+$ and \textbf{(b)} K$^+$ concentration.}
    \label{fig:si:Cu100_LiK}
\end{figure}

\subsection{Dependence of activation barriers, reaction energies, and density weighted orientations on mean work function}

\begin{figure}[H]
    \centering
    \includegraphics[width=0.55\linewidth]{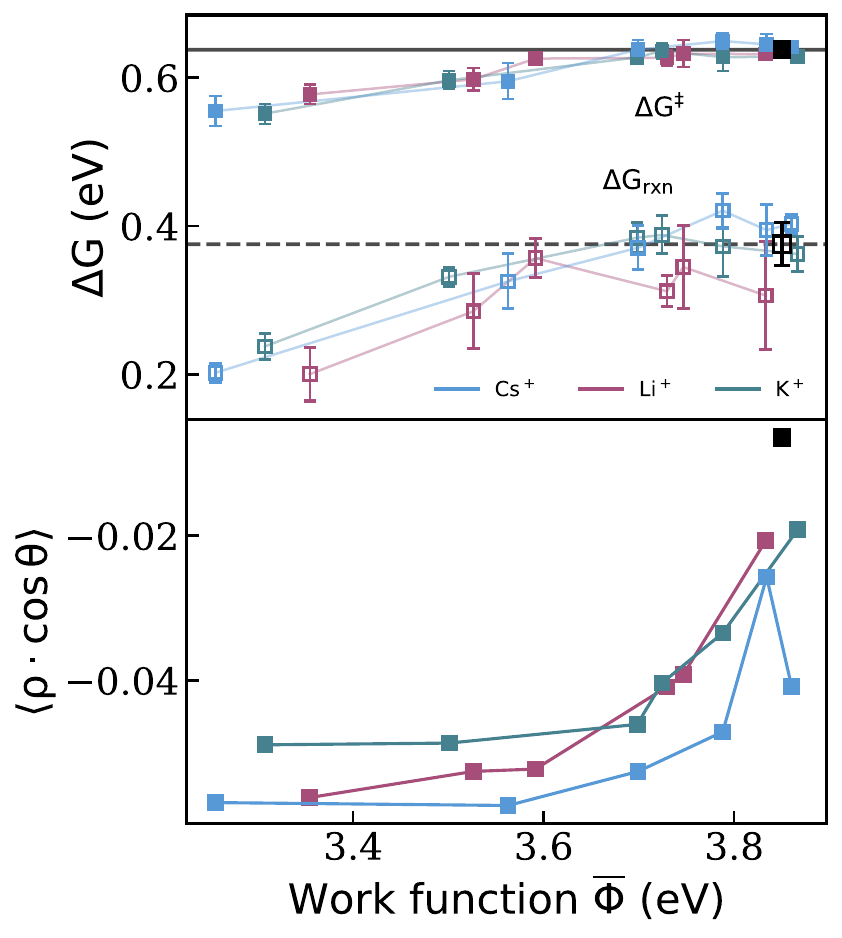}
    \caption{(top) Extracted activation barriers (filled) and reaction energies (open) from explicitly enhanced sampling across cations Cs$^+$, K$^+$, and Li$^+$ for cell size 8$\times$8 as a function of mean work function. (bottom) Density weighted orientations as a function of mean work function.}
    \label{fig:si:wf_dependence}
\end{figure}

\newpage
\subsection{Work function dependence on ion type and surface charge density}
\begin{table}[ht]
\centering
\caption{Work functions $\Phi$ (eV) across ion type, count, and frame along the reaction coordinate. For each entry, workfunctions are computed from an average of $\sim$100 samples of that frame. The aggregate workfunction is computed across all frames.}
\label{tab:workfunction}
\begin{tabular}{lcccc|c}
\toprule
 & & \multicolumn{4}{c}{$\Phi$ [eV]} \\
\cmidrule(lr){3-6}
Ion & Count & Initial & TS & Final & Aggregate \\
\midrule
\multirow{1}{*}{-} & 0 & 3.92 $\pm$ 0.08 & 3.82 $\pm$ 0.08 & 3.81 $\pm$ 0.08 & 3.85 $\pm$ 0.08 \\
\midrule
\multirow{6}{*}{Cs$^{+}$} & 2 & 3.80 $\pm$ 0.07 & 3.83 $\pm$ 0.07 & 3.87 $\pm$ 0.08 & 3.83 $\pm$ 0.08 \\
 & 4 & 3.87 $\pm$ 0.08 & 3.81 $\pm$ 0.09 & 3.90 $\pm$ 0.08 & 3.86 $\pm$ 0.08 \\
 & 5 & 3.76 $\pm$ 0.08 & 3.82 $\pm$ 0.08 & 3.79 $\pm$ 0.07 & 3.79 $\pm$ 0.08 \\
 & 6 & 3.70 $\pm$ 0.08 & 3.73 $\pm$ 0.07 & 3.66 $\pm$ 0.08 & 3.70 $\pm$ 0.08 \\
 & 7 & 3.50 $\pm$ 0.08 & 3.53 $\pm$ 0.09 & 3.65 $\pm$ 0.08 & 3.56 $\pm$ 0.08 \\
 & 8 & 3.23 $\pm$ 0.11 & 3.21 $\pm$ 0.09 & 3.33 $\pm$ 0.09 & 3.26 $\pm$ 0.09 \\
\midrule
\multirow{6}{*}{K$^{+}$} & 2 & 3.82 $\pm$ 0.08 & 3.86 $\pm$ 0.08 & 3.92 $\pm$ 0.07 & 3.87 $\pm$ 0.08 \\
 & 4 & 3.85 $\pm$ 0.08 & 3.73 $\pm$ 0.08 & 3.78 $\pm$ 0.07 & 3.79 $\pm$ 0.08 \\
 & 5 & 3.69 $\pm$ 0.08 & 3.71 $\pm$ 0.09 & 3.77 $\pm$ 0.09 & 3.72 $\pm$ 0.08 \\
 & 6 & 3.65 $\pm$ 0.08 & 3.71 $\pm$ 0.08 & 3.74 $\pm$ 0.08 & 3.70 $\pm$ 0.08 \\
 & 7 & 3.50 $\pm$ 0.08 & 3.52 $\pm$ 0.08 & 3.48 $\pm$ 0.08 & 3.50 $\pm$ 0.08 \\
 & 8 & 3.22 $\pm$ 0.08 & 3.31 $\pm$ 0.08 & 3.40 $\pm$ 0.08 & 3.31 $\pm$ 0.08 \\
\midrule
\multirow{6}{*}{Li$^{+}$} & 2 & 3.79 $\pm$ 0.07 & 3.84 $\pm$ 0.08 & 3.87 $\pm$ 0.08 & 3.83 $\pm$ 0.08 \\
 & 4 & 3.75 $\pm$ 0.07 & 3.74 $\pm$ 0.08 & 3.75 $\pm$ 0.08 & 3.75 $\pm$ 0.08 \\
 & 5 & 3.66 $\pm$ 0.08 & 3.72 $\pm$ 0.09 & 3.80 $\pm$ 0.06 & 3.73 $\pm$ 0.08 \\
 & 6 & 3.54 $\pm$ 0.08 & 3.59 $\pm$ 0.08 & 3.65 $\pm$ 0.07 & 3.59 $\pm$ 0.08 \\
 & 7 & 3.45 $\pm$ 0.07 & 3.51 $\pm$ 0.08 & 3.63 $\pm$ 0.11 & 3.53 $\pm$ 0.09 \\
 & 8 & 3.21 $\pm$ 0.07 & 3.32 $\pm$ 0.08 & 3.54 $\pm$ 0.08 & 3.35 $\pm$ 0.08 \\
\bottomrule
\end{tabular}
\end{table}

\subsection{Validation of MLIP-predicted energies and forces via DFT}
\begin{table}[H]
\centering
\begin{tabular}{@{}lcccccc@{}}
\toprule
Ion & Count
& \shortstack{Energy (meV)}
& \shortstack{Corrected Energy (meV)}
& \shortstack{Forces (meV/Å)}
& \shortstack{$\Delta G^{\ddagger}$ (meV)}
& \shortstack{$\Delta G_{\mathrm{rxn}}$ (meV)}\\ \midrule
-                   & 0       &   80.11    &  80.11   &    4.06    &   92.6    &    46.8    \\ 
\midrule
\multirow{3}{*}{Cs} & 2       &   75.48    &  71.1    &    3.83    &   99.5    &    57.5     \\
                    & 5       &   295.6    &  44.91   &    3.61    &   85.5    &    43.1     \\
                    & 8       &   450.52   &  70.75   &    4.38    &   34.3    &    41.4     \\ 
\midrule
\multirow{3}{*}{K}  & 2       &   60.95    &  51.3    &    3.99    &   78.6    &    56.1      \\
                    & 5       &   209.83   &  38.9    &    4.03    &   86.8    &    46.2      \\
                    & 8       &   308.53   &  51.03   &    4.82    &   47.8    &    47.1      \\ 
\midrule
\multirow{3}{*}{Li} & 2       &   59.04    &  63.12   &    3.99    &   97.6    &    56.6      \\
                    & 5       &   253.66   &  44.96   &    3.98    &   82.5    &    36.3      \\
                    & 8       &   314.22   &  68.29   &    4.61    &   47.4    &    31.1      \\ 
\bottomrule
                    
\end{tabular}
\caption{MLIP errors over the sampled reaction coordinate for the CO dimerization on 8$\times$8 Cu(100). Energy and force errors are evaluated as the mean absolute error (MAE) across all configurations. Energy error after applying per-ion corrections are also presented. $\Delta G^{\ddagger}$ and  $\Delta G_{\mathrm{rxn}}$ errors are computed based on the average energy of the respective state. }
\label{si:tab:ml_vs_dft}
\end{table}

\section{CO dimerization on Cu(310)}

\begin{figure}[H]
    \centering
    \includegraphics[width=0.55\linewidth]{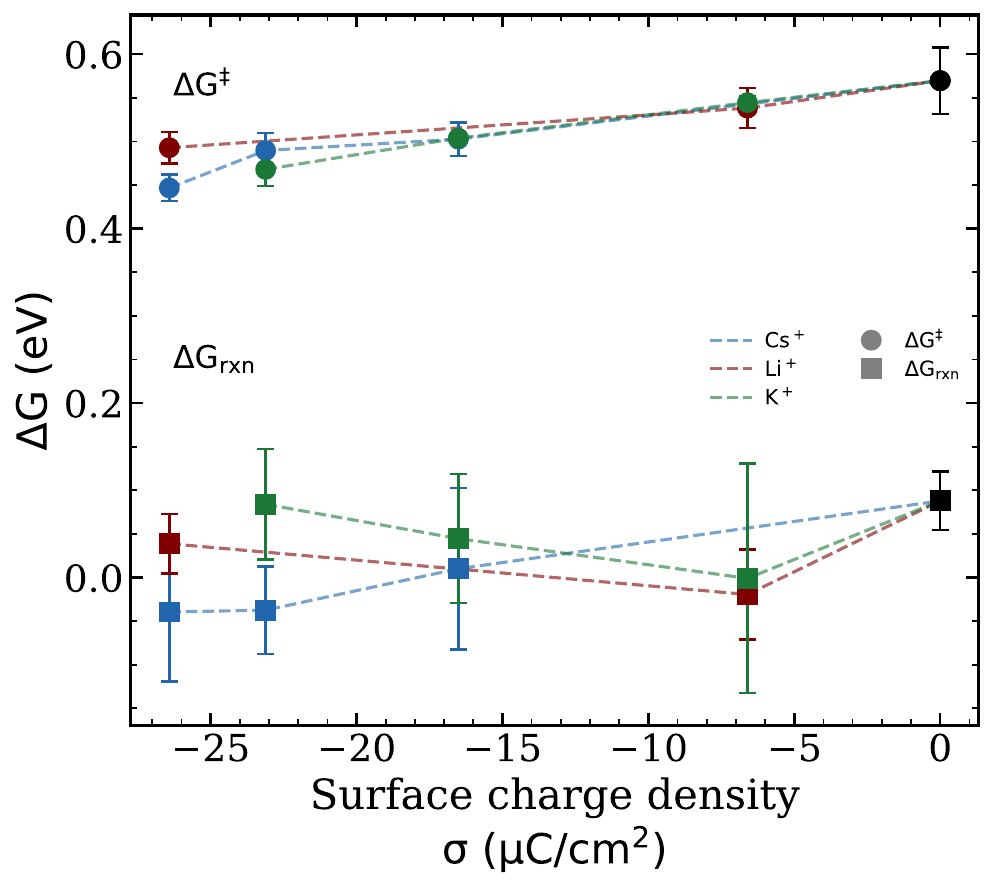}
    \caption{Extracted activation barriers (circle) and reaction energies (square) from explicitly enhanced sampling across cations Cs$^+$, K$^+$, and Li$^+$ as a function of surface charge density.}
    \label{fig:si:Cu310_barriers_energies}
\end{figure}

\end{document}